\documentclass[a4paper,11pt]{article}
\pdfoutput=1
\usepackage[utf8]{inputenc}
\usepackage[pdftex]{graphicx}
\usepackage{amsfonts,amsmath,amssymb,textcomp}
\usepackage{setspace}

\usepackage[top=2.7cm,left=2.7cm,right=2.7cm,bottom=2.7cm]{geometry}
\usepackage{color}
\usepackage[usenames,dvipsnames]{xcolor}
\usepackage{notoccite}
\usepackage{subcaption}
\usepackage{jheppub}
\usepackage{tikz}
\usetikzlibrary{3d}
\graphicspath{{figures/}}

\usepackage{hyperref}
\usepackage{float}
\usepackage{bm}
\usepackage{soul}

\DeclareMathAlphabet{\mathpzc}{OT1}{pzc}{m}{it}
\numberwithin{equation}{section}

\newcommand{\be}{\begin{equation}}
\newcommand{\ee}{\end{equation}}
\newcommand{\bea}{\begin{eqnarray}}
\newcommand{\eea}{\end{eqnarray}}

\subheader{\begin{flushright}UTTG-07-19
\end{flushright}}

\title{Constraining higher order gravities with subregion duality}
\author[a]{Elena C\'aceres,}
\author[a]{Anderson Seigo Misobuchi,}
\author[b]{Juan F. Pedraza}
\affiliation[a]{Theory Group, Department of Physics, University of Texas, Austin, TX 78712, USA}
\affiliation[b]{Institute for Theoretical Physics, University of Amsterdam,
Amsterdam, 1090 GL, NL}
\emailAdd{elenac@utexas.edu}
\emailAdd{anderson.misobuchi@utexas.edu}
\emailAdd{jpedraza@uva.nl}
\abstract{In higher derivative theories, gravity can travel slower or faster than light.
With this feature in mind, we revisit the construction of the causal and entanglement wedges in this type of theories, and  argue that they must be constructed using the fastest mode instead of null rays. We show that the property of causal wedge inclusion, i.e., the fact that the causal wedge must be contained in the entanglement wedge, leads to more stringent constraints on the  couplings than those imposed by hyperbolicity and boundary causality.
Our results imply that the full power of subregion-subregion duality
could lead to the same conclusions previously obtained based on high energy graviton scattering. We illustrate our findings with a systematic analysis in Gauss-Bonnet gravity.}

\date{\today}

\begin{document}

\maketitle

\section{Introduction}

Recent research efforts in the joint program on quantum information and holography indicate   a deep connection between spacetime and quantum entanglement. The starting point  of these developments was the groundbreaking discovery that, for quantum field theories that are holographically dual to Einstein gravity, the entanglement entropy of a region $\mathcal{A}$ can be computed as the area of a codimension-two bulk extremal surface $\Gamma_\mathcal{A}$,
\be\label{HRT}
S_\mathcal{A}=\frac{\text{Area}(\Gamma_\mathcal{A})}{4G_N}\,,
\ee
anchored at the boundary, with $\partial \Gamma_\mathcal{A}=\partial \mathcal{A}$. This relation was first conjectured in \cite{Ryu:2006bv,Hubeny:2007xt} and later proved in \cite{Lewkowycz:2013nqa,Dong:2016hjy} using the basic ingredients of the holographic dictionary.

\begin{figure}[t!]
 \centering
 \begin{center}
 $\qquad\quad$
   \includegraphics[width=0.3\linewidth]{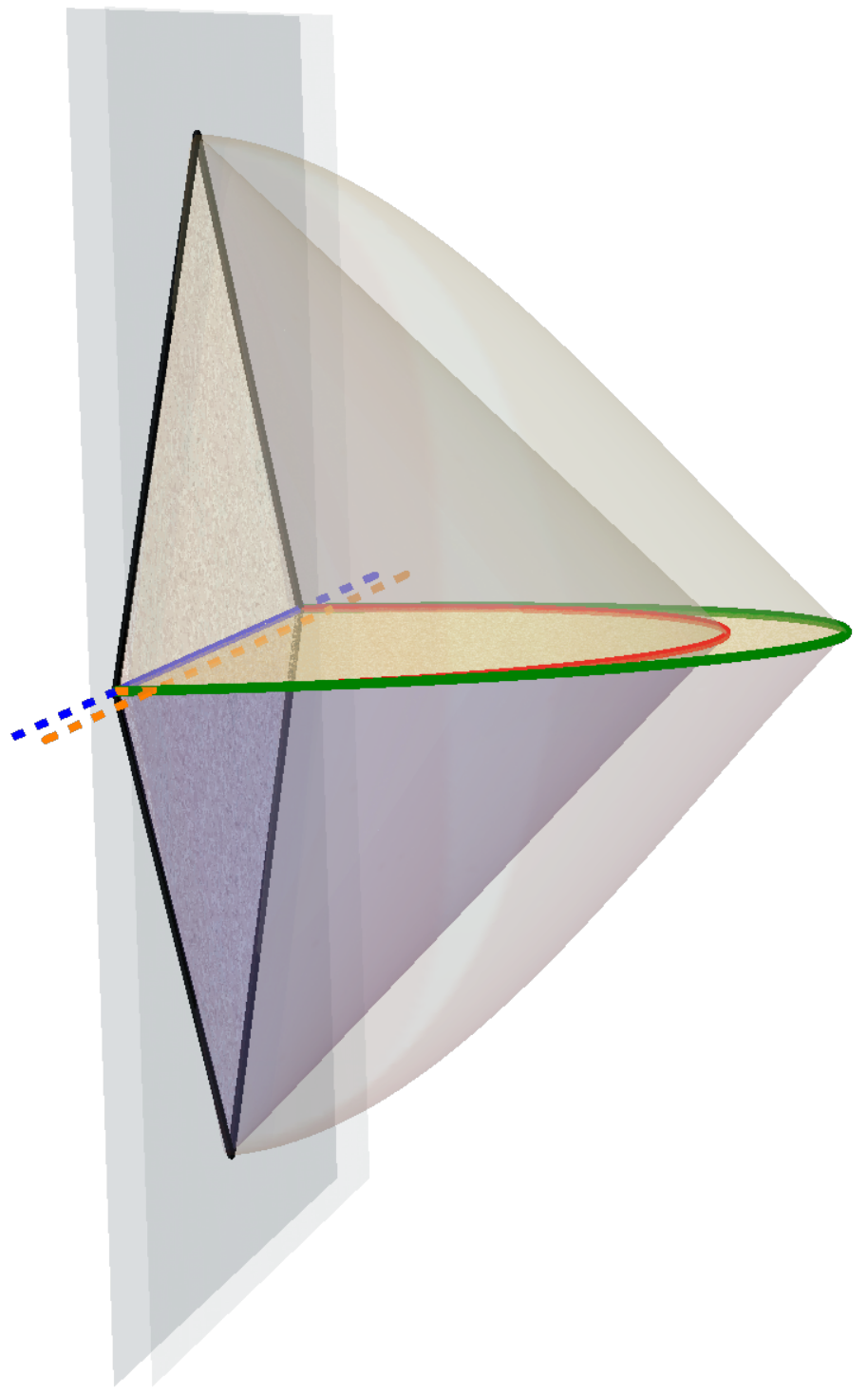}
\begin{picture}(0,0)
\put(-5,112){{\scriptsize $\Gamma_\mathcal{A}$}}
\put(-46,112){{\scriptsize $\Xi_\mathcal{A}$}}
\put(-42,150){{\scriptsize $\mathcal{E}(\mathcal{A})$}}
\put(-89,150){{\scriptsize $\mathcal{C}(\mathcal{A})$}}
\put(-105,114){{\scriptsize $\mathcal{A}$}}
\put(-110.5,130){{\scriptsize $\mathcal{D}[\mathcal{A}]$}}
\put(-138,10){{\scriptsize $z=0$}}
\put(-97,10){{\scriptsize $z=\delta$}}
\end{picture}
\begin{tikzpicture}[x  = {(0.9cm,0cm)},
                    y  = {(0.5cm,0.5cm)},
                    z  = {(0cm,0.9cm)},
                    scale = 1,
                    color = {lightgray}]
\hspace{-2.3cm}
\begin{scope}[canvas is xy plane at z=3]
  \draw[black,->] (0,0) -- (1,0) node (x-axis) [right] {{\scriptsize$\!z$}};
  \draw[black,->] (0,0) -- (0,1) node (y-axis) [above] {{\scriptsize$\,\,\,\,\,x$}};
\end{scope}
\begin{scope}[canvas is yz plane at x=0]
  \draw[black,->] (0,3) -- (0,4) node (z-axis) [above] {{\scriptsize$t$}};
\end{scope}

\end{tikzpicture}
\end{center}
 \vspace{-6mm}
 \caption{Schematic representation of the entanglement wedge $\mathcal{E}(\mathcal{A})$ and causal wedge $\mathcal{C}(\mathcal{A})$ associated with a boundary region $\mathcal{A}$, or equivalently, with its boundary domain of dependence $\mathcal{D}[\mathcal{A}]$. The entanglement wedge is the bulk region bounded by null geodesics that are shot towards the boundary from the extremal surface $\Gamma_\mathcal{A}$. For general bulk geometries, $\mathcal{E}(\mathcal{A})$ contains the causal wedge $\mathcal{C}(\mathcal{A})$, which is the bulk region bounded by null geodesics that are shot into the bulk from the edge of $\mathcal{D}[\mathcal{A}]$.
 The spatial surface $\Xi_A$ on which the latter geodesics intersect is the bifurcation surface of the causal wedge. In a few situations, $\Xi_\mathcal{A}=\Gamma_\mathcal{A}$ and the two wedges coincide, but in general cases, $\Gamma_\mathcal{A}$ reaches deeper into the bulk. This property has been rigorously shown to hold for generic bulk geometries in Einstein gravity but the proof does not apply for higher order derivative gravities. However, consistency of subregion-subregion duality implies that it should be true for any theory with a physically sensible holographic dual.}
 \label{fig:wedges}
\end{figure}
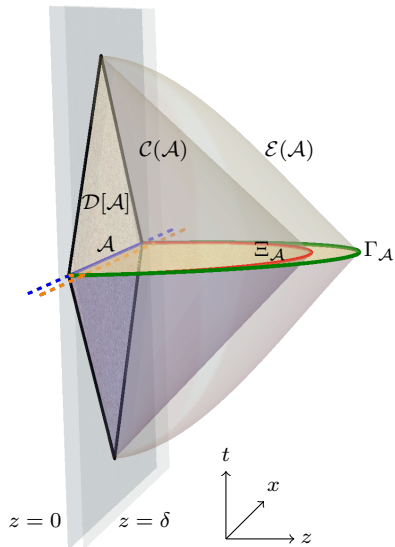

In quantum field theory, the entanglement entropy of a region, $\mathcal{A}$, and its complement, $\mathcal{A}^c$, is defined as the von Neumann entropy, $S_\mathcal{A}\equiv -\text{Tr}(\rho_\mathcal{A}\log\rho_\mathcal{A})$. The quantity $\rho_\mathcal{A}=\text{Tr}_{\mathcal{A}^c}(\rho)$ is the reduced density matrix associated with $\mathcal{A}$, or more specifically, with  its  domain  of  dependence,  $\mathcal{D}[\mathcal{A}]$. In the holographic context, relation (\ref{HRT}) led to the notion of subregion-subregion duality \cite{VanRaamsdonk:2009ar,Czech:2012bh,Bousso:2012sj,Bousso:2012mh,Headrick:2014cta} and  to the idea that the quantum state of the boundary theory reduced to a subregion $\mathcal{A}$ (encoded in its reduced density matrix $\rho_\mathcal{A}$) is dual to a particular subregion in the bulk, the entanglement wedge $\mathcal{E}(\mathcal{A})$. The entanglement wedge is defined as the domain of dependence of any codimension-one bulk spacelike surface bounded by $\Gamma_\mathcal{A}$ and $\mathcal{A}$. See Figure \ref{fig:wedges} for a more detailed explanation. Consistency of subregion duality has led to numerous insights in the context of gauge/gravity duality, ranging from the problem of bulk reconstruction, quantum error correction and tensor networks constructs of AdS/CFT \cite{Almheiri:2014lwa,Pastawski:2015qua,Hayden:2016cfa,Dong:2016eik,Harlow:2016vwg,Donnelly:2016qqt,Faulkner:2017vdd,Cotler:2017erl,Espindola:2017jil,Espindola:2018ozt,Faulkner:2018faa}. At the same time, the duality also imposes several constraints on the bulk geometry. For an excellent account of these constraints and the connections among them see \cite{Akers:2016ugt}.

Another important construct is the causal wedge $\mathcal{C}(\mathcal{A})$, defined as the bulk region $I^-(\mathcal{D}[\mathcal{A}])\cap I^+(\mathcal{D}[\mathcal{A}])$\footnote{$I^\pm$ denote the chronological future/past regions in the bulk of the boundary domain of dependence $D[\mathcal{A}]$ of a region $\mathcal{A}$.}.
The causal wedge in a spacetime that has a CFT dual has to be  completely contained in the entanglement wedge $\mathcal{C}(\mathcal{A})\subset\mathcal{E}(\mathcal{A})$ \cite{Wall:2012uf}; see Figure \ref{fig:wedges} for a schematic illustration. We will refer to this property as Causal Wedge Inclusion (CWI).  The bifurcation surface of the causal wedge, denoted as $\Xi_\mathcal{A}$, plays a special role in its interpretation. In Einstein gravity, it is generically contained within $\Gamma_\mathcal{A}$ and has more area than the latter. Moreover, since $\Xi_\mathcal{A}$ can be defined covariantly, its area was proposed in \cite{Hubeny:2012wa} as a coarse-grained measure of entanglement, dubbed the \emph{causal holographic information}. This work led to a number of proposals regarding the precise information-theoretic quantity in the boundary theory that is dual to its area \cite{Freivogel:2013zta,Kelly:2013aja} that were later shown not to be viable \cite{Engelhardt:2017wgc}. More recently, in \cite{Engelhardt:2017lfb} the authors proposed a concrete recipe for coarse-graining in the dual quantum field theory that is consistent with the holographic constructs. This  led to the definition of a new object, called the \emph{reduced causal density  matrix}, which was argued to be the natural candidate for the dual of the causal wedge. Regardless of its precise information-theoretic interpretation, the important point here is that by consistency of subregion-subregion duality, any field theory observable defined within $\mathcal{D}[\mathcal{A}]$ should be accessible from the region of the bulk geometry that is dual to the actual \emph{reduced density  matrix}, i.e. the entanglement wedge $\mathcal{E}(\mathcal{A})$. Hence, CWI \emph{must} hold true for \emph{all} gravity theories with a physically sensible holographic dual.

Consistency of AdS/CFT also demands the  Boundary Causality Condition (BCC), which is the condition that the causal structure of the boundary theory is compatible with the causal structure of the bulk. More specifically, the statement of BCC implies that two boundary points that are null related through the boundary theory point of view cannot be connected by a timelike curve through the bulk. In Einstein gravity, it can be shown that CWI $\Rightarrow $ BCC \cite{Akers:2016ugt}. However, in higher derivative theories of gravity, the above relation does not hold generically. In these theories there are several subtleties to consider, so one must analyze case by case. On one hand, the holographic prescription to compute entanglement entropy includes non-universal corrections involving extrinsic curvature terms \cite{Dong:2013qoa,Camps:2013zua}, so the surface $\Gamma_A$ minimizes a theory-dependent functional. On the other hand, in higher derivative theories the characteristic hypersurfaces are generically \emph{not} null, which implies that gravity can propagate slower or faster than the speed of light \cite{Izumi:2014loa,Reall:2014pwa}. If the latter is true, this could in particular lead to possible violations of boundary causality.

In theories with superluminal propagation of gravitons in the bulk, the causal structure is no longer determined by null rays since the fastest mode can probe larger regions in the bulk. One crucial consequence, in the context of subregion duality, is that the definition of the causal and entanglement wedges must be adapted to account for these modes. The causal wedge $\mathcal{C}(\mathcal{A})$ is usually defined as the bulk region bounded by a family of null geodesics that are shot into the bulk from the edge of $\mathcal{D}[\mathcal{A}]$. In a theory with superluminal modes, the geodesics that we shoot must follow the fastest mode of propagation instead of null rays. Similarly, the entanglement wedge $\mathcal{E}(\mathcal{A})$ is generated by shooting null geodesics from the extremal surface $\Gamma_\mathcal{A}$ towards the boundary, thus for a theory with superluminal propagation we need to replace these null geodesics by geodesics of the fastest mode.

In this paper, we argue that these new ingredients, together with the CWI property, drastically  constrain the space of parameters of couplings in general higher order gravities. The bulk action of these theories can be generically written as follows
\be
S=\frac{1}{16\pi G_N}\int d^{d+1}\sqrt{-g}\left[R-2\Lambda+\alpha_1\left(R^2\text{ terms}\right)+\alpha_2\left(R^3\text{ terms}\right)+\cdots\right]\,,
\ee
where $\Lambda=-\frac{d(d-1)}{2L^2}$. If we consider this theory of gravity as a low-energy effective theory with a UV cutoff of order $M_P$, then the natural values of the higher derivative couplings are of order $\alpha_i\sim\mathcal{O}(\ell_P^{2i})$. Couplings of these orders cannot be ruled out. Alternatively, one could take these couplings to be arbitrary, e.g. $\alpha_i\gg \ell_P^{2i}$, but then it is not clear if these theories would make sense in the UV (without the need of arbitrarily higher order terms in the expansion). Our analysis lead to constrains for such large couplings.

For concreteness we focus on Gauss-Bonnet gravity, where the Einstein-Hilbert term is only supplemented by certain combination of curvature squared terms. However, our methodology could be easily adapted to constrain generic higher order gravitational theories. In this case, we normalize the extra coupling as $\alpha_1\equiv\lambda L^2$, so that we would be concerned with values of $\lambda$ of order $\mathcal{O}(1)$. The specific solutions that we will consider throughout this paper are spherical AdS black holes in Gauss-Bonnet gravity, which are dual to thermal states in the boundary theory. The physics of the spherical solutions is richer than their planar counterparts;
due to finite volume effects, the entanglement entropy generically undergoes a first order phase transition \cite{Hubeny:2013gta} and exhibits shadows \cite{Freivogel:2014lja}. Moreover, the causal wedges can display nontrivial topology \cite{Headrick:2014cta}, which makes the analysis of the CWI property more intricate. We will also test if CWI $\Rightarrow$ BCC holds in our higher derivative setup. The strategy here is to test the negative, namely that nBCC $\Rightarrow$ nCWI, so that we one can start from cases where BCC is violated \cite{Andrade:2016yzc}.

We want both CWI and BCC to hold  for all possible states of the theory.
One can argue that the set of \textit{all states} includes small black holes, regardless of their thermodynamics, because they are part of the Hilbert space of the theory. In fact, even though small black holes evaporate, they survive for a very long time compared to their energy. This suggests that they are approximately metastable and should be considered in the microcanonical ensemble;\footnote{
  The thermodynamics of Gauss-Bonnet black holes has been extensively studied and computed using different methods \cite{Cho:2002hq,Astefanesei:2008wz,Liu:2008zf}. By focusing on a specific ensemble and requiring the thermodynamic stability of these solutions, one could get additional constraints on the GB coupling. However, for the purpose of deriving a the state-independent constraint we will not impose such restriction here, nor restrict our attention to a particular ensemble.
  }
see \cite{Asplund:2008xd,Hanada:2016pwv,Berenstein:2018lrm,Marolf:2018ldl} for proposals addressing the form of the corresponding dual states from the boundary theory perspective.

We find that the two conditions above significantly constrain the value of the higher derivative couplings.  For Gauss-Bonnet gravity, we find that
\emph{any} $\mathcal{O}(1)$ value of the Gauss-Bonnet parameter $\lambda$ leads to a violation of CWI and is therefore inconsistent.
This agrees with the conclusion of \cite{Camanho:2014apa} based on positivity bounds from high energy graviton scattering.
If we restrict ourselves to large black holes, the window of allowed couplings almost closes but is still finite.
An interesting possibility to consider in the future is that other subregion properties like entanglement wedge nesting and possibly quantum corrections, might be enough to rule out any value of the Gauss-Bonnet coupling for \emph{any} size black hole.

The paper is organized as follows. In section \ref{sec:2}, we start with some background material required for 
the subsequent sections. We begin by explaining the definition of characteristic hypersurfaces and reviewing their construction in generic higher order gravities of the Lovelock type. Then, we specialize to Gauss-Bonnet gravity and present basic aspects of the spherical black hole solutions of the theory. We give explicit expressions for the speed of propagation of different modes of the theory and show the implications for boundary causality and hyperbolicity. In section \ref{sec:3} we present explicit calculations of entanglement entropy in these gravity solutions and study the behavior of the extremal surfaces $\Gamma_\mathcal{A}$ as we vary the size of the region, the black hole mass and the Gauss-Bonnet coupling. We give particular attention to the study of the entanglement shadows. In section \ref{sec:4} we construct the causal surfaces $\Xi_\mathcal{A}$ and study their dependence on the aforementioned parameters. We proceed to test the CWI property and perform a detailed analysis of the cases where it is violated.
We close in section \ref{sec:5} with some final remarks and possible future directions.

\section{Background\label{sec:2}}

\subsection{Characteristic hypersurfaces}

In general relativity, when faced with a system of partial differential equations obtained from Einstein's equations, oftentimes  we want to choose
initial data on a  Cauchy slice and evolve this initial data in time. Generically, it might happen that we find a  hypersurface beyond which  the evolution of the differential equations is not unique. This hypersurface is called ``characteristic hypersurface". Thus, identifying the characteristic hypersurfaces allows us to study the causal structure of a system of PDEs.

In Einstein gravity, the characteristic surfaces are the null surfaces of the spacetime, so if the initial data is defined on  a spacelike surface,  we are guaranteed that the evolution of  this initial data is unique.  But this is not the case for more general theories of gravity; in higher derivative theories some degrees of freedom may be spacelike with respect to the propagation of light rays. In this paper, we are interested in investigating certain holographic constructs in theories where the characteristic hypersurfaces are not null.

In \cite{Reall:2014pwa, Izumi:2014loa} the authors studied the characteristic hypersurfaces for tensor, vector, and scalar graviton degrees of freedom in spherically symmetric black hole solutions of Lovelock theories. Below, following  closely  \cite{Andrade:2016yzc}, we  summarize some results of \cite{Reall:2014pwa, Izumi:2014loa} that we will use in the rest of the paper.

Consider a $d+1$ dimensional spacetime with coordinates $x^\mu$ with metric which we denote as $g_I$. In Lovelock theories the equations of motion $\mathcal{E}_J=0$ depend linearly on $\partial_0^2 g_I$. Thus $\mathcal{E}_J$ can be written as
\begin{equation}\label{eq:eom}
\frac{\partial \mathcal{E}_J}{\partial(\partial^2_0 g_{I})} \partial_0^2 g_{I} + \dots =0.
\end{equation}
If we know $g_{I}$ and its derivatives in a given hypersurface $\Sigma$ with coordinates $(x_0=0, x_i)$ then, just by acting with $\partial_i$, we also know $\partial _i\partial_\mu g_I$ . However, $\partial_0^2 g_{I} $ has to be determined from the equations of motion \eqref{eq:eom}, which will only have unique solutions if the matrix
$$\frac{\partial \mathcal{E}_J}{\partial(\partial^2_0 g_{I})}, $$
also called the principal symbol, is invertible. In this case $\Sigma$ is non-characteristic and we can uniquely evolve the equations of motion starting with initial data on $\Sigma$. If the principal symbol is not invertible, $\Sigma$ is characteristic, the equations are not hyperbolic and the initial value problem with $\Sigma$ as the starting surface is ill-posed.

If we consider metric fluctuations, it can be shown that the fastest mode propagates along a characteristic hypersurface.
To see this consider solving the equations of motion imposing initial conditions on a  hypersurface $\mathpzc{A}$. The edge of the Cauchy development of  $\mathpzc{A}$ must be  described by the fastest propagation mode. On the other hand, the characteristic hypersurface is a boundary beyond which the dynamical equations of motion cannot be uniquely solved, \emph{i.e.} the characteristic hypersurface is the edge of the Cauchy development. Thus, the fastest modes propagate along a characteristic hypersurface.

A familiar example is Einstein gravity; it can be shown that the characteristic initial data problem is well defined. Moreover, characteristic hypersurfaces in GR are always null, and thus gravity travels at the speed of light.

This is not the case in Gauss-Bonnet (GB) gravity \cite{Reall:2014pwa, Izumi:2014loa}. However,  in symmetric spacetimes, we can define an effective metric  such that the characteristic hypersurface for a given degree of freedom is null with respect to that metric.

In black hole solutions of the form
\begin{equation}\label{eq:physical}
ds^2= -f(r) dt^2 +\frac{dr^2}{f(r)} + r^2 d\Omega_{d-2}^2,
\end{equation}
where $d\Omega^2_{d-2}$ is the line element of a  space $S$, linear perturbations can be classified as scalar, vector or tensor with respect to the symmetries of $S$. For each type of perturbation the equations of motion will lead to a master equation that can be written as a wave equation with a potential
\begin{equation}\label{eq:master}
\left ( -\frac{\partial^2}{\partial t^2} + \frac{\partial^2}{\partial r_*^2} - V^l_{A}(r)\right ) \Psi^l_{A}(t,r)=0
\end{equation}
where $A$ denotes the type of fluctuation, $A \in { T,V,S}$ and $l$ labels the harmonic.
To determine the principal symbol we need to identify  terms that involve second derivatives. We can do this by focussing on highly oscillatory modes since in this case the second derivatives will dominate the equation. For large $l$ and denoting as $D^2$ the Laplacian on $S$,  we can recast \eqref{eq:master} as
\begin{equation}
\left( -\frac{\partial^2}{\partial t^2} + \frac{\partial^2}{\partial r_*^2} - f(r)\frac{c_A (r)}{r^2} D^2 \right ) \Psi_A(t,r)=f(r) G_A^{\mu\nu} \partial_\mu\partial_\nu \Psi_A=0.
\end{equation}

Thus, for each mode the  characteristic hypersurface is null with respect to the corresponding effective metric:
\begin{equation}\label{eq:effective}
G_{A \mu\nu} dx^\mu dx^\nu = - f(r) dt^2 + \frac{dr^2}{f(r)} + \frac{r^2}{c_A (r)} d\Omega^2.
\end{equation}
In a $d+1$ Lovelock theory we have
\begin{align}
	c_T(r)&=-\left( 1+\frac{1}{d-3}\right) A(r) -\left( 1-\frac{1}{d-3}\right)\frac{1}{A(r)} +B(r)+3\\
	c_V(r)&=A(r)\\
	c_S(r)&=3\left( 1-\frac{1}{d-1}\right)A(r)+\left( 1-\frac{3}{d-1}\right)\frac{1}{A(r)}-\left( 1-\frac{2}{d-1}\right)(B(r)+3).
	\end{align}
The functions $A(r)$ and $B(r)$ for a general Lovelock theory  are known but are cumbersome and not enlightening so we will not reproduce them here.
The null cones of $G_{A \mu \nu} $ determine causality of the theory in the physical spacetime \eqref{eq:physical}.
Note that \cite{Brigante:2008gz} the factor $\frac{f(r) c_A (r)}{r^2} $ can be interpreted as the local speed of the perturbation on a constant $r$ hypersurface.

In a Lovelock theory the characteristic determinant factorizes as a product of the effective metrics
\begin{equation}\label{eq:factorQ}
Q(x,\xi)=( G_S^{ a b} (x)\xi_a \xi_b)^{n_S}( G_V^{ a b}(x)\xi_a \xi_b)^{n_V} ( G_T^{a b}(x)\xi_a \xi_b)^{n_T},
\end{equation}
where $n_S, n_V, n_T$ are the numbers of degrees of freedom of the corresponding perturbation. From \eqref{eq:factorQ} we see that if any of the $c_A$ vanishes for some $r>r_H$, then there is no surface providing good initial data for that solution; that is, the solution is not hyperbolic.

\subsection{Gauss-Bonnet gravity}

Lovelock gravity \cite{Lovelock:1971yv} is a generalization of Einstein gravity including higher curvature terms in the Lagrangian. The equivalent of the Einstein tensor in this class of theories satisfies the properties of being symmetric, divergence-free, and it depends only on the metric and its first and second derivatives. The simplest example of a Lovelock theory that includes a higher curvature term is given by Gauss-Bonnet gravity in $d+1>4$ spacetime dimensions,\footnote{In $d+1\leq4$ the Gauss-Bonnet term is topological so it does not affect the bulk equations of motion.} whose action for $d=4$ is
\begin{equation}
 I=\frac{1}{16\pi G_N}\int d^{5}x\sqrt{-g}\left[R+\frac{12}{L^2}+\frac{\lambda
L^2}{2}(R^2-4R_{\mu\nu}R^{\mu\nu}+R_{\mu\nu\rho\sigma}R^{\mu\nu\rho\sigma})\right],
\end{equation}
where $L$ is a length scale and $\lambda$ is the Gauss-Bonnet coupling constant. This theory admits black hole solutions which are asymptotically AdS \cite{Nojiri:2001aj,Cai:2001dz}. We will focus on the 5-dimensional AdS black hole solution with spherical horizon described by the metric
\begin{gather} \label{eq:metric}
 ds^2=-\frac{f(r)}{f_\infty}dt^2+\frac{dr^2}{f(r)}+r^2(d\phi^2+\sin^2\phi\,d\Omega^2_2),\\
 f(r)=1+\frac{r^2}{2\lambda L^2}\left(1-\sqrt{1-4\lambda+4\lambda\frac{\mu}{r^4}}\right),\\
 f_\infty=\frac{1-\sqrt{1-4\lambda}}{2\lambda},\quad \mu=r_h^4+r_h^2L^2+\lambda L^4,
\end{gather}
where $d\Omega^2_2=d\chi^2+\sin^2\omega \,d\omega^2$ is the metric of a unit 2-sphere. The factor
$f_{\infty}$ is chosen such that the speed of light at the boundary ($r\to\infty$) is one, and $\mu$ is a parameter related to the ADM mass of the black hole. The solution also needs to obey the condition
\begin{equation} \label{eq:negbound}
 r_h^2+2\lambda L^2>0,
\end{equation}
which is necessary to ensure that $f(r_h)=0$. The solution is asymptotically AdS with radius dependent on $\lambda$ given by $L_{AdS}=L/\sqrt{f_\infty}$.

\subsubsection{Boundary causality and hyperbolicity}

\begin{figure}[t!]
    \centering
    \begin{subfigure}[]{0.42\textwidth}
        \includegraphics[width=\textwidth]{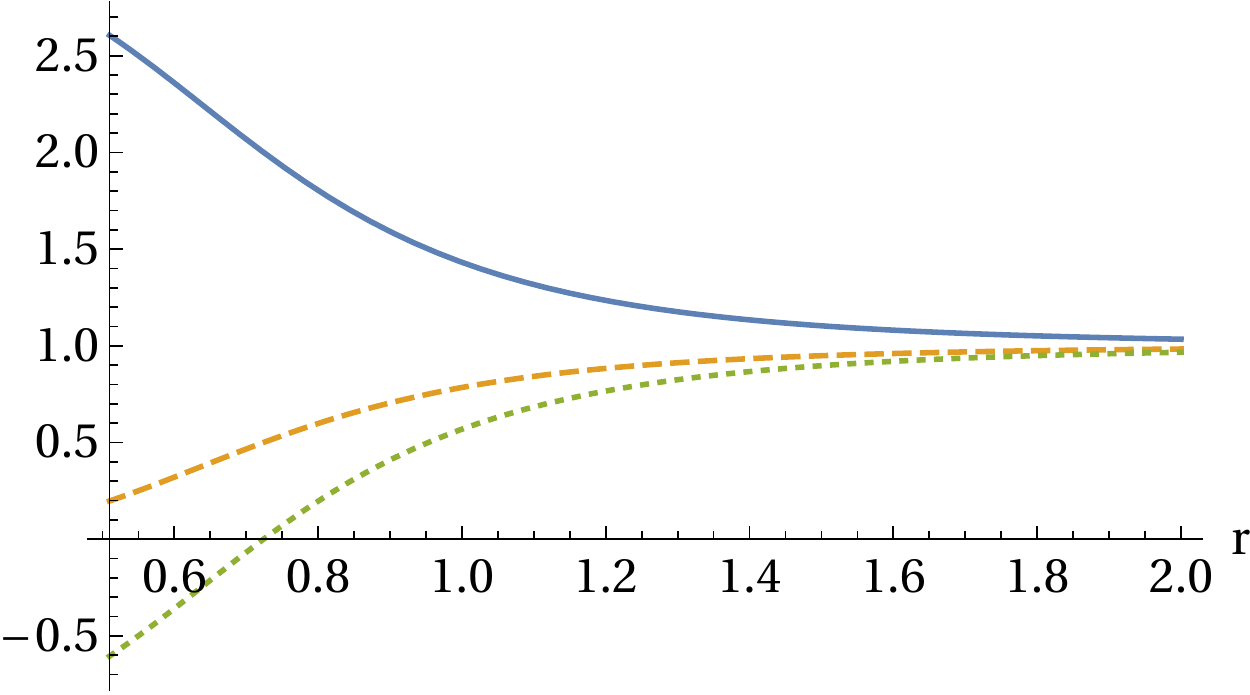}
        \caption{$\lambda>0$}
        \label{fig:cpositive}
    \end{subfigure}
    \begin{subfigure}[]{0.52\textwidth}
        \includegraphics[width=\textwidth]{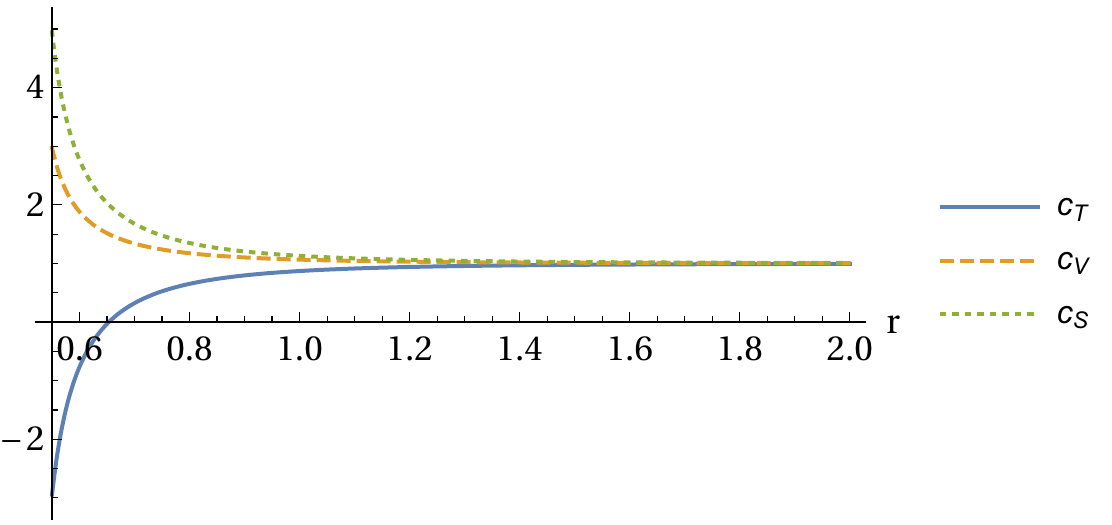}
        \caption{$\lambda<0$}
        \label{fig:cnegative}
    \end{subfigure}
     \vspace{-2mm}
    \caption{Speed of propagation for tensor, vector and scalar modes. For $\lambda>0$, we have $c_T>c_V>c_S$ and $c_T>c_V>0$ for all $r$. For $\lambda<0$, we have $c_S>c_V>c_T$ and $c_S>c_V>0$ for all $r$. (a) $r_h=0.5,\,\lambda=0.1$. (b) $r_h=0.5,\,\lambda=-0.1$.
    }\label{fig:c}
\end{figure}

Combining results from the analysis of the propagation of scalar, vector and tensor modes for metric fluctuations in Gauss-Bonnet spacetimes \cite{Dotti:2005sq,Gleiser:2005ra}, the causal structure of the background \eqref{eq:metric} can be studied by using the effective metric \cite{Reall:2014pwa}
\begin{equation} \label{eq:metriceff}
 ds^2=-\frac{f(r)}{f_\infty}dt^2+\frac{dr^2}{f(r)}+\frac{r^2}{c_A(r)}d\Omega_3^2,
\end{equation}
where $c_A$ (the index $A=T,V,S$ labels tensor, vector, and scalar, respectively) is the speed of propagation of the mode. Null geodesics in the effective metrics \eqref{eq:metriceff} correspond to geodesics for the superluminal or subluminal modes in the original metric \eqref{eq:metric}. Explicitly, the speeds of propagation for each mode are given by

\begin{equation}\label{eq:cAs}
 \begin{cases}
  c_T(r)=-2g(r)+3\\
  c_V(r)=g(r)\\
  c_S(r)=2g(r)-1
 \end{cases}, \qquad
 g(r)=\frac{(1-4\lambda)r^4}{2\lambda (r_h^4+r_h^2L^2+\lambda L^4)+(1-4\lambda)r^4}.
\end{equation}
From \eqref{eq:cAs} one can show that in a GB  theory with a positive coupling,  $c_T>c_V>c_S$ for all $r$ and thus the tensor modes travel fastest.
However,  $c_S$ is the only one that can become negative, so it is the scalar mode that determines if the equations of motion are hyperbolic around a given solution.
Conversely, if the  coupling is negative, the fastest modes are the scalars and it is the tensors which are related to hyperbolicity violations. In Figure \ref{fig:c} we plot \eqref{eq:cAs} for different values of the coupling.

Therefore, hyperbolicity  violation is determined by the roots of $c_T$, $c_V$ and $c_S$ but the causal structure is determined by the mode that travels fastest.  Consistent solutions will preserve boundary causality as well as hyperbolicity. This analysis was carried out in \cite{Andrade:2016yzc} and the authors found that for small black holes hyperbolicity is a more stringent constraint than causality.
In the present work, we want to  investigate the  inclusion of the causal wedge in the entanglement wedge as an additional requirement, that we will show, is even more constraining than causality and hyperbolicity.

\begin{figure}[t!]
    \centering
    \begin{subfigure}[]{0.4\textwidth}
        \includegraphics[width=\textwidth]{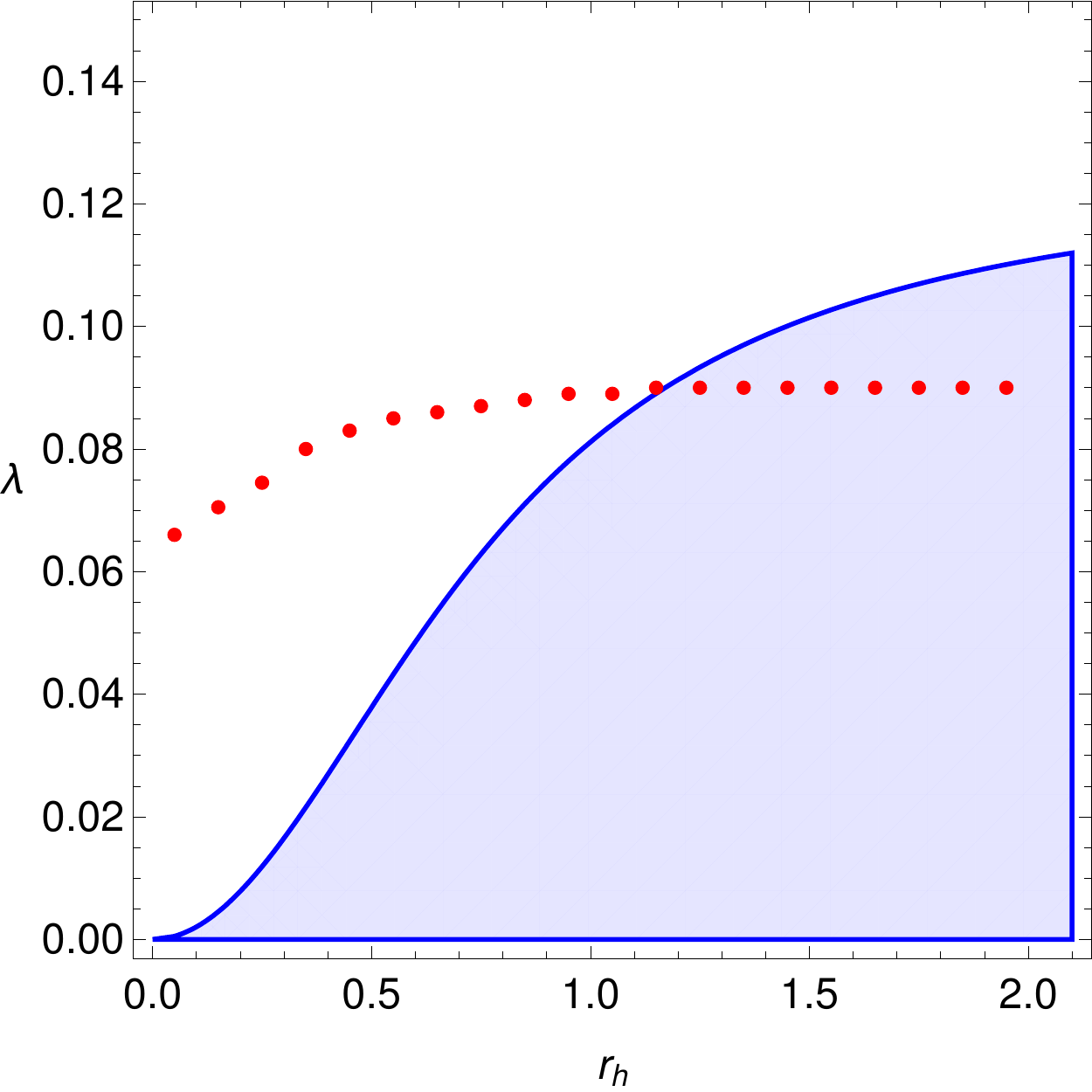}
        \label{fig:constraints0_lambdapos}
    \end{subfigure}
    \qquad
    \begin{subfigure}[]{0.4\textwidth}
        \includegraphics[width=\textwidth]{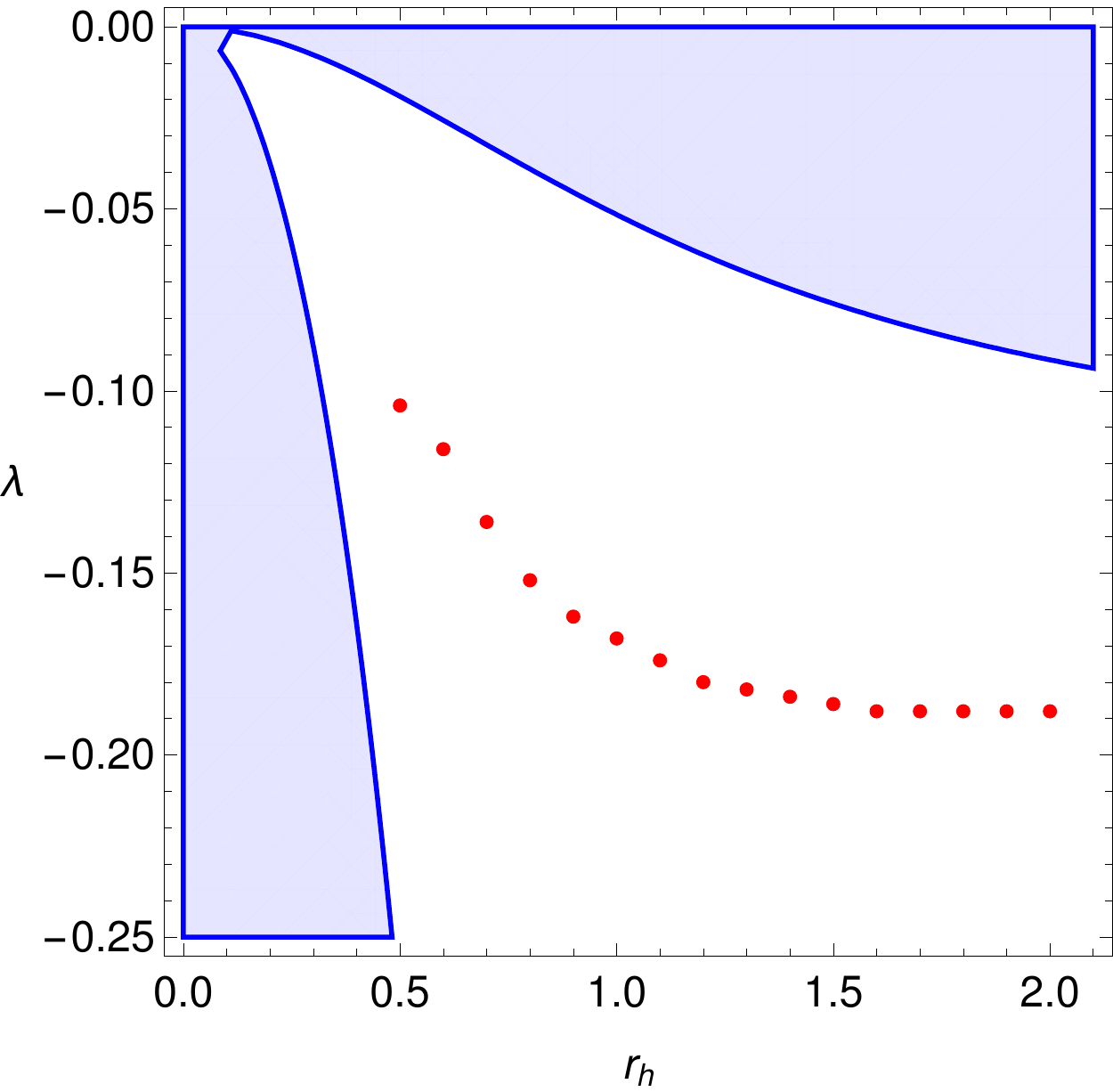}
        \label{fig:constraints0_lambdaneg}
    \end{subfigure}
     \vspace{-6mm}
    \caption{
Results from \cite{Andrade:2016yzc}. Constraints on the parameter space $(r_h,\lambda)$ of the spherical AdS Gauss-Bonnet black hole soluton due to: hyperbolicity (Blue), and boundary causality (Red). The region outside the shaded region or above (below in the right figure) the dots is not allowed.
    }
    \label{fig:constraints0}
\end{figure}

For the reference, in Figure \ref{fig:constraints0} we repeat the result of \cite{Andrade:2016yzc} summarizing the constraints from boundary causality and hyperbolicity.\footnote{If we consider further higher derivative terms, several new constraints could be studied, e.g. \cite{Takahashi:2011du}.}

\section{Holographic entanglement entropy in Gauss-Bonnet\label{sec:3}}

Let us first discuss the recipe to compute entanglement entropy in higher order gravities. After the implementation of the replica trick in the bulk it follows that, for a $(d+1)$-dimensional gravity theory with Lagrangian built from arbitrary contractions of Riemann tensors $\mathcal{L}(R_{\mu\nu\rho\lambda})$, the entanglement entropy functional (\ref{HRT}) generalizes to \cite{Dong:2013qoa,Camps:2013zua}
\be\label{HRT-General}
S_{\mathcal{A}}= 2\pi\int_{\Gamma_{\mathcal {A}}}\!\!\! d^{d-1} \sigma \sqrt{h} \left\{ \frac{\partial \mathcal{L}}{\partial R_{z \bar{z} z \bar{z}}} + \sum_\alpha \left(\frac{\partial^2 \mathcal{L}}{\partial R_{zizj} \partial R_{\bar{z} k \bar{z} l}}\right)_\alpha \frac{8K_{zij} K_{\bar{z} kl}}{q_\alpha+1} \right\} \,.
\ee
In this formula $h$ denotes the determinant of the induced metric on the surface $\Gamma_\mathcal{A}$, $z$ and $\bar{z}$ are complex coordinates transverse to the surface, and $K_{zij}$ and $K_{\bar{z} kl}$ are extrinsic curvatures. Finally, in the second derivative of $\mathcal{L}$ one expands in terms of the extrinsic curvature $K_{aij}$, $Q_{abij} \equiv \partial_a K_{bij}$. For each term (labeled by $\alpha$) one defines $q_\alpha$ as the number of $Q_{zzij}$ and $Q_{\bar{z}\bar{z} ij}$, plus one half of the number of $K_{aij}$, $R_{abci}$, and $R_{aijk}$. The first term above is identical to the Wald entropy functional, while the second term is an anomaly-like contribution that only arises in theories of quadratic or higher order in the curvature. For Einstein gravity only the first term contributes and gives rise to the HRT formula (\ref{HRT}).

\subsection{Entanglement surfaces}

Let us now specialize to Gauss-Bonnet gravity. In this case, the corrected functional that gives the entanglement entropy reduces to \cite{Hung:2011xb, deBoer:2011wk}
\begin{equation} \label{eq:functional}
 S_\mathcal{A}=\frac{1}{4G_N}\int_{\Gamma_\mathcal{A}}\!\!\! d^{3}\sigma\sqrt{h}(1+\lambda L^2 \mathcal{R})+\frac{\lambda
L^2}{2G_N}\int_{\partial\Gamma_\mathcal{A}}\!\!\!d^2\sigma\sqrt{\tilde{h}}\, \mathcal{K},
\end{equation}
The second term here is a boundary term needed to make the variational principle well-defined. Again, $h$ and $\tilde{h}$ are the induced metric determinants
on $\Gamma_\mathcal{A}$ and $\partial \Gamma_{\mathcal{A}}$, respectively. $\mathcal{R}$ is the scalar curvature on $\Gamma_\mathcal{A}$
and $\mathcal{K}$ is the trace of the extrinsic curvature at $\partial \Gamma_{\mathcal{A}}$. Since the geometry of our interest is static, we fix the $t=0$ slice and take the boundary region $\mathcal{A}$ to be a disk centered at the north pole bounded by the hypersurface $\phi=\phi_\mathcal{A}$ ($0\leq \phi_\mathcal{A}\leq \pi$). In the limit $\phi_\mathcal{A}\to\pi$, the region $\mathcal{A}$ covers the entire boundary space.
Specializing the functional \eqref{eq:functional} to the background \eqref{eq:metric}, we obtain (see the Appendix \ref{sec:functional} for derivation)
\begin{equation} \label{eq:functional2}
 S_\mathcal{A}=\frac{\pi}{G_N}\int d\sigma\left[\sqrt{h_{\sigma\sigma}}\left(r^2\sin^2\phi+2\lambda
L^2\right)+\frac{2\lambda L^2}{\sqrt{h_{\sigma\sigma}}}\left(r\phi'\cos\phi +
r'\sin\phi\right)^2\right],
\end{equation}
where $r=r(\sigma),\phi=\phi(\sigma)$, prime denotes derivative w.r.t. the parameter $\sigma$,\footnote{By choosing $\phi$ as the parameter $\sigma$, the boundary conditions are $r(\phi_\mathcal{A})=r_\text{max}$ (radial boundary cutoff) and $r'(0)=0$
(surface smooth at $\phi=0$). However, the parametrization $r(\phi)$ fails when we consider large enough boundary regions because the derivative $r'(\phi)$ diverges at some intermediate point.  For numerical purposes, we choose the alternative parametrization $\phi(r)$ to evolve the remaining part of the equation until we reach the boundary.} and
\begin{equation}
 h_{\sigma\sigma}(\sigma)=\frac{{r'}^2(\sigma)}{f(r(\sigma))}+r^2(\sigma){\phi'}^2(\sigma).
\end{equation}
\begin{figure}[t!]
    \centering
    \begin{subfigure}[]{0.38\textwidth}
        \includegraphics[width=\textwidth]{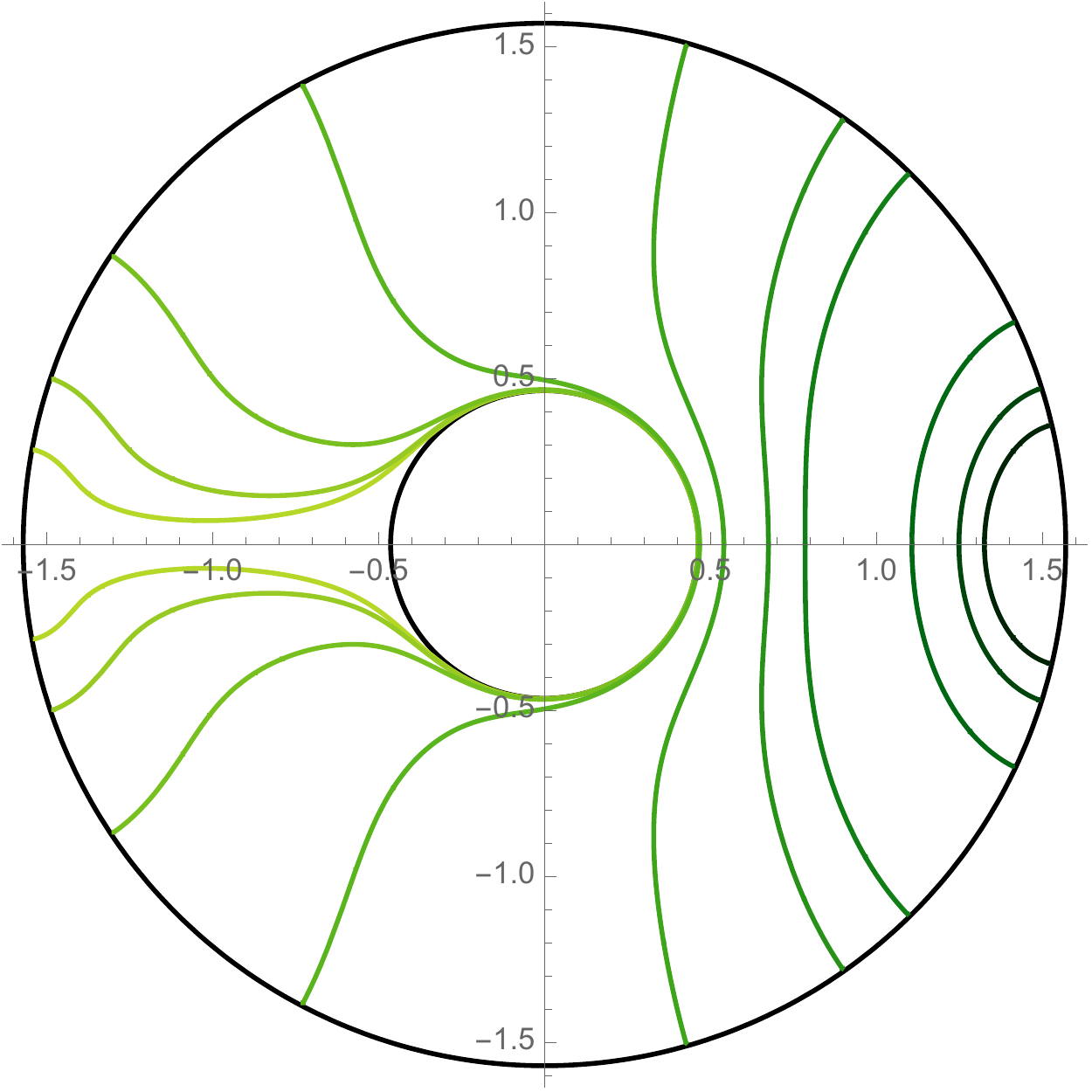}
        \caption{$\lambda>0$}
        \label{fig:ew_lambdapos}
    \end{subfigure}
    \qquad
    \begin{subfigure}[]{0.38\textwidth}
        \includegraphics[width=\textwidth]{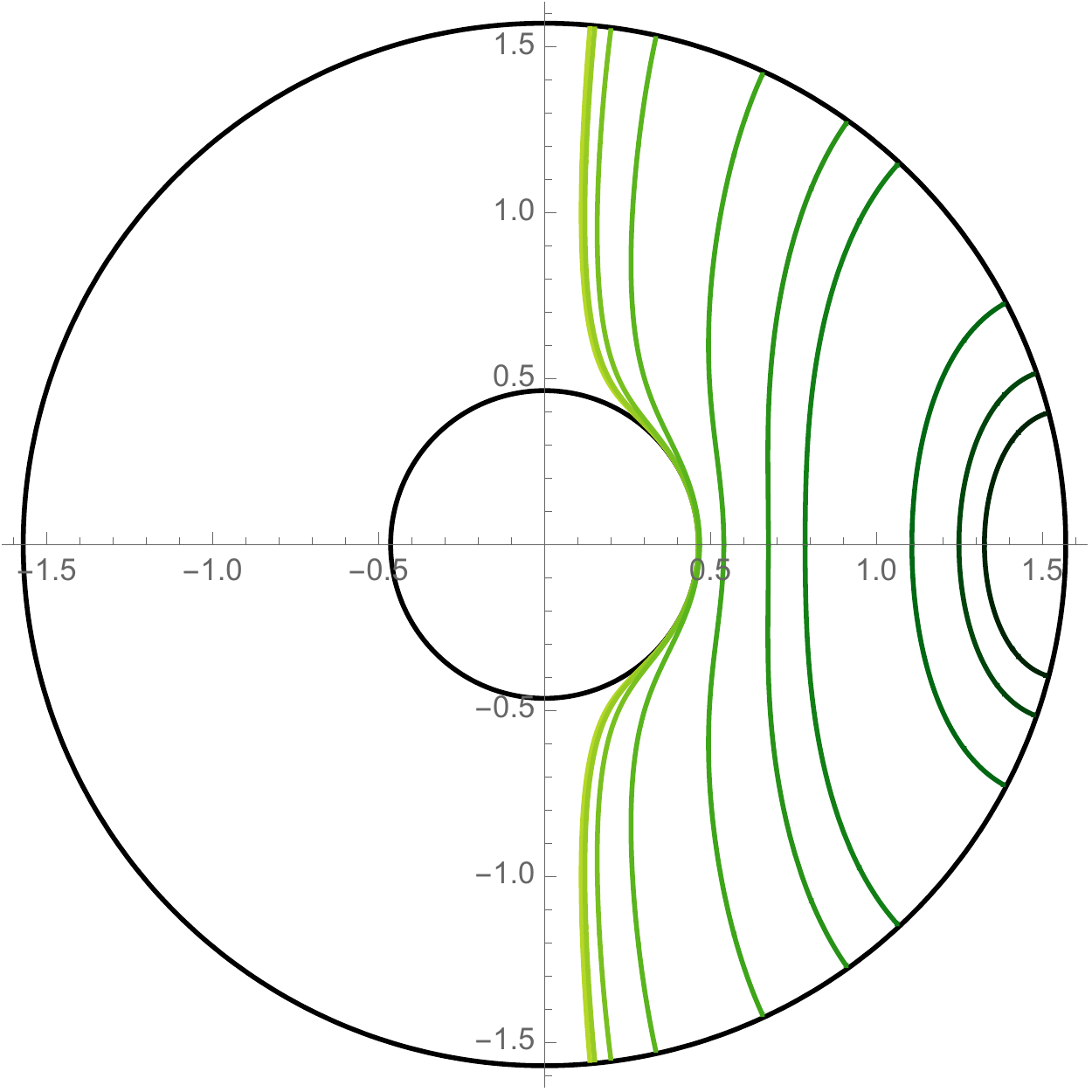}
        \caption{$\lambda<0$}
        \label{fig:ew_lambdaneg}
    \end{subfigure}
     \vspace{-2mm}
    \caption{Entanglement minimal surfaces $\Gamma_\mathcal{A}$ for the functional \eqref{eq:functional2}, at fixed horizon radius $r_h=0.5$ and fixed Gauss-Bonnet coupling $\lambda$. We adopt polar coordinates $(\rho,\phi)$, where $\rho=\tan^{-1}r$ is a compactified radial coordinate. The inner and outer circles represent the black hole horizon and the boundary, respectively. The angle $\phi_\mathcal{A}$ that defines the entangling surface has range $0\leq \phi_\mathcal{A}\leq\pi$, so that the curve that appears at the bottom is just a reflection $\phi\to-\phi$. (a) Fixed $\lambda=0.1$ (b) Fixed $\lambda=-0.1$. We chose the same values of $r_0$ (minimal radius) for both left and right plots, namely, $r_0=0.50015, 0.5002, 0.5005, 0.505, 0.6, 0.8, 1, 2, 3, 4$ (from lighter to darker green).}
    \label{fig:ew_lambda}
\end{figure}

\begin{figure}[t!]
    \centering
    \begin{subfigure}[]{0.35\linewidth}
        \includegraphics[width=\linewidth]{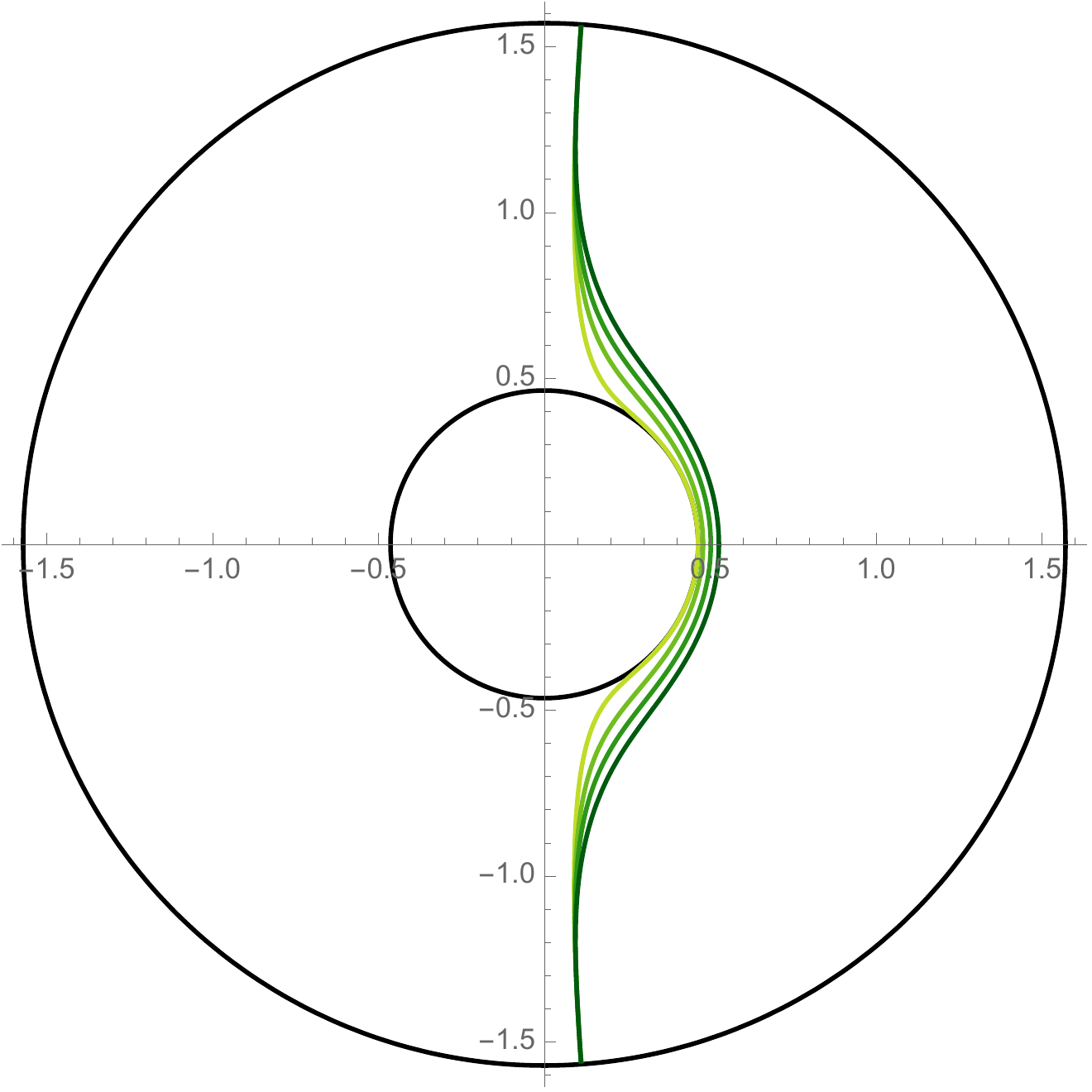}
        \caption{}
        \label{fig:ew1}
    \end{subfigure}
    \ \
    \begin{subfigure}[]{0.52\linewidth}
        \includegraphics[width=\linewidth]{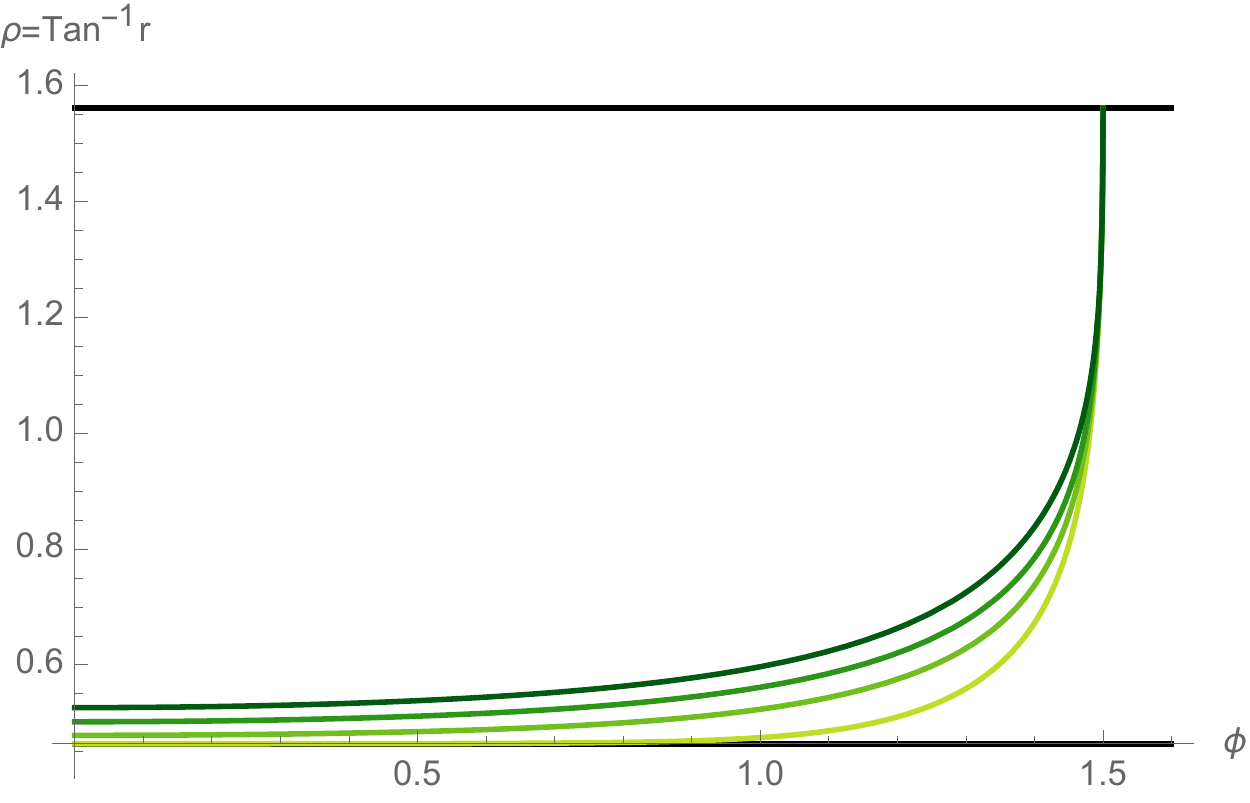}
        \caption{}
        \label{fig:ew2}
    \end{subfigure}
     \vspace{-2mm}
    \caption{Entanglement minimal surfaces $\Gamma_\mathcal{A}$ for fixed boundary region $\mathcal{A}$ of size parametrized by the polar angle $\phi_\mathcal{A}$. We have set $r_h=0.5$ and $\phi_\mathcal{A}=1.5$. (a) From lighter to darker green: $\lambda= -0.1, 0, 0.1, 0.2$. Plot in (b) is just a cartesian version of (a).}
    \label{fig:ew}
\end{figure}

We solve the equations starting from some minimal radius $r_0$, so that $r_0$ can be adjusted to give the correct value of $\phi_\mathcal{A}$. Figure \ref{fig:ew_lambda} shows a family of minimal surfaces for different sizes of the boundary spatial region and fixed Gauss-Bonnet coupling $\lambda$. The numerical solution is plotted in polar coordinates $(\rho,\phi)$, where $\rho\equiv\tan^{-1}r$ is a compactified radial coordinate. For $\lambda>0$, the minimal surface reaches larger values of $\phi_\mathcal{A}$, while the opposite happens for $\lambda<0$. In Figure \ref{fig:ew}, the boundary region is kept fixed and we notice that the minimal surfaces becomes closer to the horizon as we decrease $\lambda$.

The entanglement entropy is obtained by evaluating the functional \eqref{eq:functional} on the minimal surface. There are two candidates for the minimal surface: the minimal surface homologous to $\mathcal{A}$ and the minimal surface homologous to $\mathcal{A}^c$ plus the black hole horizon surface. We expect a phase transition characterizing the stage at which these two candidates exchange dominance \cite{Hubeny:2013gta}. The transition is determined by a critical value $\phi_\mathcal{E}^*$ such that for $\phi_\mathcal{A}\geq \phi_\mathcal{E}^*$ the Araki-Lieb inequality
\begin{equation}
 |S_\mathcal{A}-S_{\mathcal{A}^c}|\leq S_{\text{thermal}}
\end{equation}
is saturated. In other words, an `entanglement plateau' should appear when we consider the ratio $|S_\mathcal{A}-S_{\mathcal{A}^c}|/S_{\text{thermal}}$ as a function of $\phi_\mathcal{A}$ for $\phi_\mathcal{A}\geq \phi_\mathcal{E}^*$. For the spherical Gauss-Bonnet black hole \eqref{eq:metric}, the thermal entropy is given by the black hole entropy \cite{Cai:2001dz}
\begin{equation}
 S_{BH}=\frac{\pi r_h^3}{2 G_N}\left(1+\frac{6\lambda L^2}{r_h^2}\right).
\end{equation}

\begin{figure}[t!]
 \centering
 \includegraphics[width=0.7\linewidth]{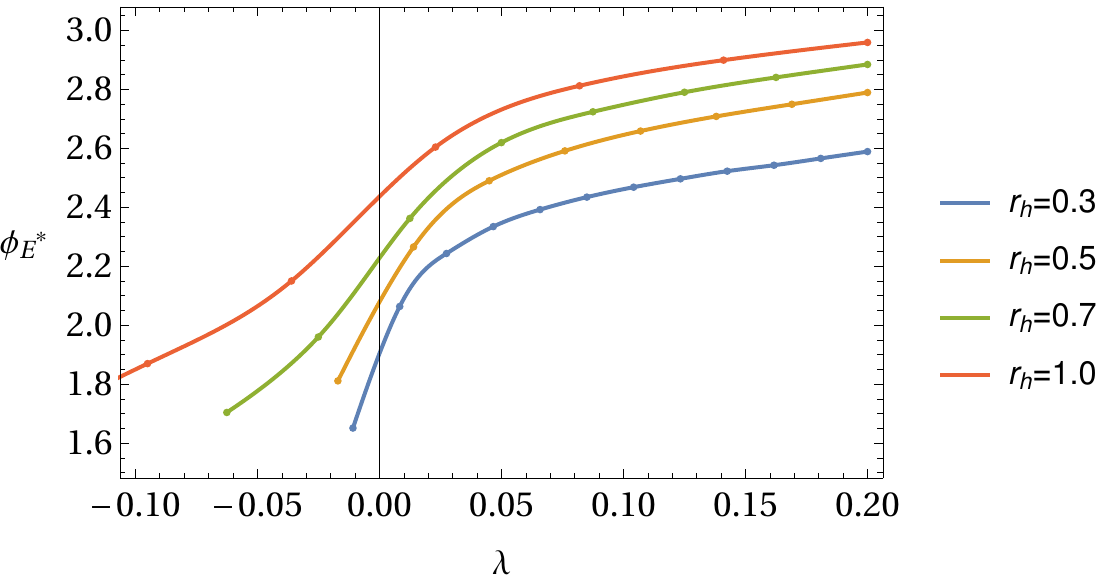}
  \vspace{-5mm}
 \caption{Critical points $\phi_\mathcal{E}^*$ for several values of horizon radius $r_h$ as a function of the Gauss-Bonnet coupling $\lambda$. The point $\phi_\mathcal{E}^*$ is defined as the value of $\phi_\mathcal{A}$ at the phase transition of entanglement entropy where the two candidates to minimal surface, one homologous to $\mathcal{A}$ and the other homologous to $\mathcal{A}^c$ plus the horizon surface, exchange dominance.}
 \label{fig:phiE}
\end{figure}

In Figure \ref{fig:phiE} we have identified the values $\phi_\mathcal{E}^*$ at which the phase transition occurs as a function of $\lambda$ for fixed values of horizon radius. Increasing $\lambda$ makes the phase transition occur at larger values of $\phi_\mathcal{A}$. For large enough negative values of $\lambda$, the phase transition becomes closer to $\pi/2$, so one might ask if there is some $\lambda$ such that there is no longer a phase transition. We checked numerically that this is not the case, essentially because \eqref{eq:negbound} places a bound on negatives values of the coupling, preventing the phase transition to disappear.

\subsection{Entanglement shadows}

For completeness, we finish this section with an investigation of the `entanglement shadow' \cite{Freivogel:2014lja}, i.e., the region in the bulk that cannot be probed by extremal surfaces. Due to the existence of a phase transition where the two candidates to minimal surface exchange dominance, there is a region surrounding the horizon that  entanglement cannot probe. This region is determined by the minimal radius $r_0$ reached by the minimal surface at the phase transition point, where the boundary region has size parametrized by $\phi_\mathcal{E}^*$. We want to investigate the dependence of the size of the shadow $\Delta r_0\equiv r_0-r_h$ as a function of the Gauss-Bonnet coupling $\lambda$.

The size of the shadow can be better estimated in the limit of small and large black holes using the solution $r(\phi)$ that minimizes the functional \eqref{eq:functional} near the horizon \cite{Freivogel:2014lja}. In this region, with the assumption that $r'(\phi)\ll 1$, the equation of motion for $r(\phi)$ drastically simplifies and can be solved analytically giving
\begin{equation}\label{eq:near}
 r(\phi)\simeq r_h+\frac{\Delta r_0\sinh(\phi\gamma)}{\gamma\sin\phi}, \quad \gamma\equiv \sqrt{\frac{5r_h^2L^2+12r_h^4-2\lambda L^4}{L^2(r_h^2+2L^2\lambda)}}
\end{equation}

In order to estimate $\Delta r_0$, we will assume that the solution is valid up to some order one factor away from the horizon. Let us take this point to be $r=2r_h$ to be more concrete. We split the analysis into large and small black holes that add more simplifications.

\paragraph{Large black hole $r_h\gg L_{AdS}$:} Far from the horizon, the minimal surface is essentially radial. Thus, up to order one constants, we can estimate the shadow by plugging $r=2r_h$ at the jointing point where the solution \eqref{eq:near} is still valid and it starts to be mostly radial, and take that $\phi=\phi_\mathcal{E}^*$. This gives

\begin{equation}
 \Delta r_0^\text{large}\simeq\frac{4\sqrt{3}r_h^2\sin\phi_\mathcal{E}^*}{L}e^{-\frac{2\sqrt{3}r_h\phi_\mathcal{E}^*}{L}}
\end{equation}

Similarly to what happens in Schwarzschild-AdS, the shadow $\Delta r_0$ is exponentially small in $r_h$ \cite{Freivogel:2014lja}, which is in agreement to the expectation that large enough black holes approach the planar limit where there is no phase transition and therefore no shadow. But for the GB solution it turns out that the exponential decay also depends on the coupling $\lambda$. From the numerical analysis of the phase transition whose results are shown in Figure \ref{fig:phiE}, we see that positive $\lambda$ makes the decay faster, while negative $\lambda$ makes the decay slower.

\paragraph{Small black hole $r_h\ll L_{AdS}$:} In this regime the contribution from the black hole horizon will be small, so we expect the phase transition to occur very close to $\pi/2$. Evaluating the solution \eqref{eq:near} again at $r=2r_h$, and assuming that $\phi=\pi/2$, the shadow is approximately given by
\begin{equation}
 \Delta r_0^\text{small}\simeq\frac{r_h\gamma}{\sinh\left(\frac{\pi}{2}\gamma\right)}, \quad \gamma=\sqrt{\frac{5r_h^2L^2-2\lambda L^4}{r_h^2+2L^2\lambda}}
\end{equation}
Note that negative values of the coupling tends to make the shadow smaller. However, the restriction \eqref{eq:negbound} imposes a lower bound on negative values of $\lambda$, preventing that the size of the shadow becomes zero.

\section{Causal wedge inclusion\label{sec:4}}

\subsection{Construction of the causal wedge}

In this section we construct the causal wedge $\mathcal{C}(\mathcal{A})$ associated to a spatial region $\mathcal{A}$ at the boundary for the Gauss-Bonnet black hole solution \eqref{eq:metric}. We take $\mathcal{A}$ to be the same spherical region bounded by the hypersuface $\phi=\phi_\mathcal{A}$ we have considered to compute the entanglement entropy minimal surface in section \ref{sec:3}. As pointed out in the Introduction, the construction of the causal wedge we propose here takes into account the gravitational superluminal modes in the bulk, so that instead of shooting null rays into the bulk we shoot geodesics following the fastest mode of propagation. For the Gauss-Bonnet black hole we consider as our toy model, the information about tensor (T), vector (V), and scalar (S) mode propagations can be nicely packaged into effective metrics \eqref{eq:metriceff} in such a way that geodesics following each mode are described by null geodesics in the effective metric.

The null geodesic equations for the effective metric \eqref{eq:metriceff} are given by
\begin{equation} \label{eq:geodesiceff}
 \dot{t}=\frac{f_\infty}{f(r)},\qquad \dot{\phi}=\frac{\ell c_A(r)}{r^2},\qquad
\dot{r}=\eta\sqrt{f_\infty-\frac{\ell^2 c_A(r)f(r)}{r^2}}, \qquad A=T,V,S
\end{equation}
where the dot denotes derivative w.r.t. some affine parameter. The sign $\eta=\pm 1$ corresponds to an ougoing/ingoing geodesic. The angular momentum $\ell$ associated to the Killing vector $\partial_\phi$ is conserved and due to reflection symmetry $\phi\to -\phi$ we can choose it to be positive $\ell\in[0,1]$.
We also define the effective potential
\begin{equation}
-V_\text{eff}^A(r)\equiv f_\infty-\frac{\ell^2 c_A(r)f(r)}{r^2}.
\end{equation}
Equations \eqref{eq:geodesiceff} can be integrated, giving
\begin{align}
 t_\text{in}(r) &= t_i +\int_{\infty}^{r}dy\frac{f_\infty}{f(y)\sqrt{-V_\text{eff}^A(y)}},\\
 t_\text{out}(r) & = t_\text{in}(r_\text{turn})+\int_{r_\text{turn}}^{r}
dy\frac{f_\infty}{f(y)\sqrt{-V_\text{eff}^A(y)}},\\
 \phi_\text{in}(r) &=-\int_{\infty}^rdy\frac{\ell c_A(y)}{{y}^2\sqrt{-V_\text{eff}^A(y)}}, \\
 \phi_\text{out}(r) & =\phi_\text{in}(r_\text{turn})+\int_{r_\text{turn}}^{r}dy\frac{\ell c_A(y)}{{y}^2\sqrt{-V_\text{eff}^A(y)}},
\end{align}
where $t_i$ is the initial condition, which we will set as $t_i=-\phi_\mathcal{A}$, the lower tip of $D[\mathcal{A}]$. The ingoing part of the solution is valid up to the turning point $r_\text{turn}$, which occurs when $\dot{r}=0$. From \eqref{eq:geodesiceff}, we notice that only geodesics with sufficiently large angular momentum $\ell\geq \ell_\text{min}$ have a turning point, where the minimum angular momentum satisfies the equation
\begin{equation} \label{eq:lmin}
 \ell_\text{min}^2\geq\frac{r^2 f_\infty}{c_A(r) f(r)}.
\end{equation}
The minimum angular momentum can be determined numerically by minimizing the right hand side of the above inequality over $r$. Once $\ell_\text{min}$ is determined, the turning point $r_\text{turn}$ can be found by solving for the equality in \eqref{eq:lmin} and taking the largest root.

The causal wedge is constructed by shooting geodesics of fastest propagation, which is the tensor mode for $\lambda>0$ and the scalar mode for $\lambda<0$, from the lower tip of $D[\mathcal{A}]$ using several values of $\ell$. Previous investigations using boundary causality to derive constraints in higher derivative theories take into account only geodesics with a turning point that starts and ends at the boundary, but for the construction of the causal wedge we also need to include geodesics that cross the horizon and end at the singularity. In general, geodesics with small angular momentum will cross the $t=0$ slice in a region closer to the horizon compared to geodesics with large angular momentum that cross the $t=0$ slice closer to the boundary. The causal information surface $\Xi_\mathcal{A}$ is then obtained by interpolating the intersections of the geodesics with different values of $\ell$ with the $t=0$ time slice.

\begin{figure}[t!]
    \centering
    \begin{subfigure}[]{0.38\textwidth}
        \includegraphics[width=\textwidth]{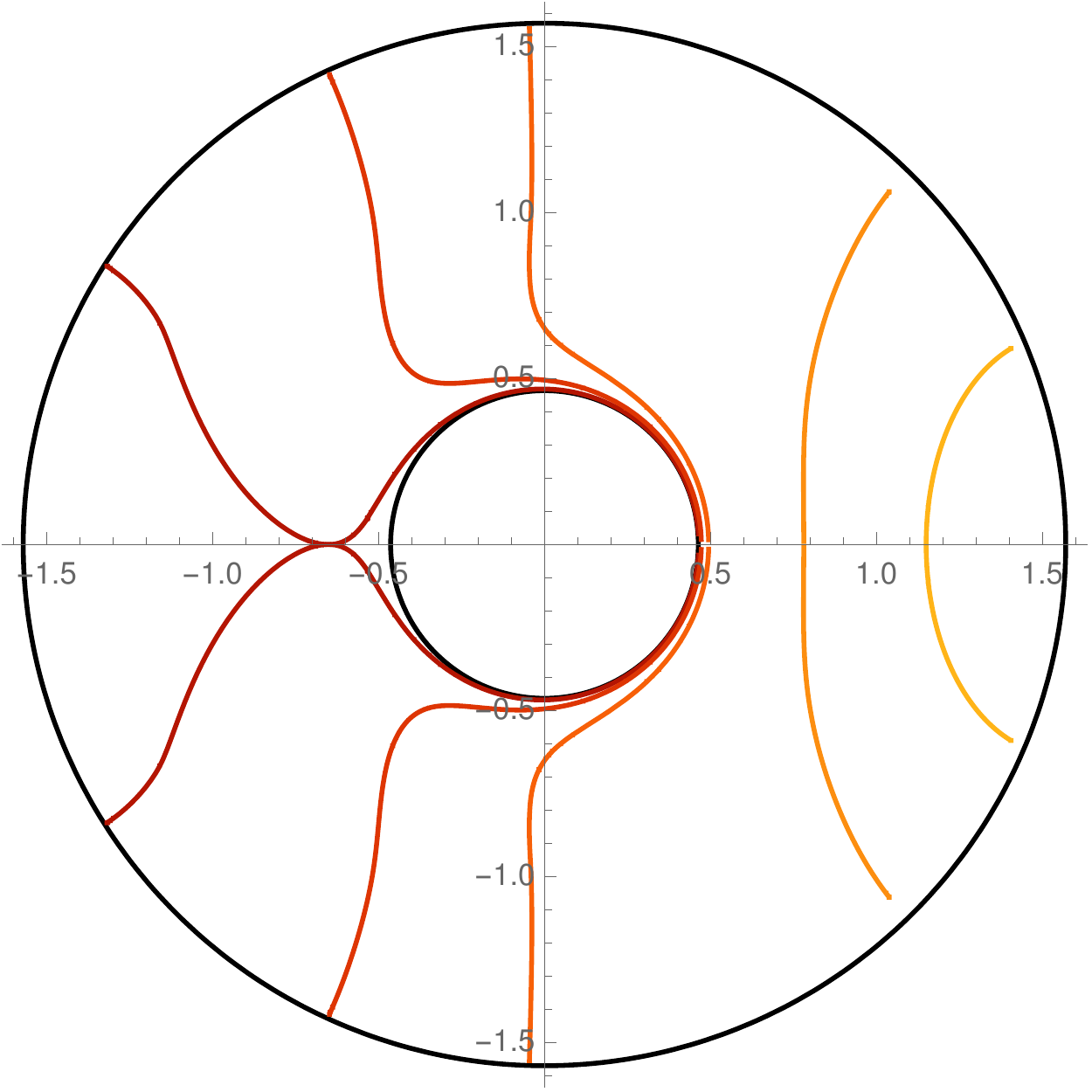}
        \caption{$\lambda>0$}
        \label{fig:fish_lambdapose}
    \end{subfigure}
    \qquad
    \begin{subfigure}[]{0.38\textwidth}
        \includegraphics[width=\textwidth]{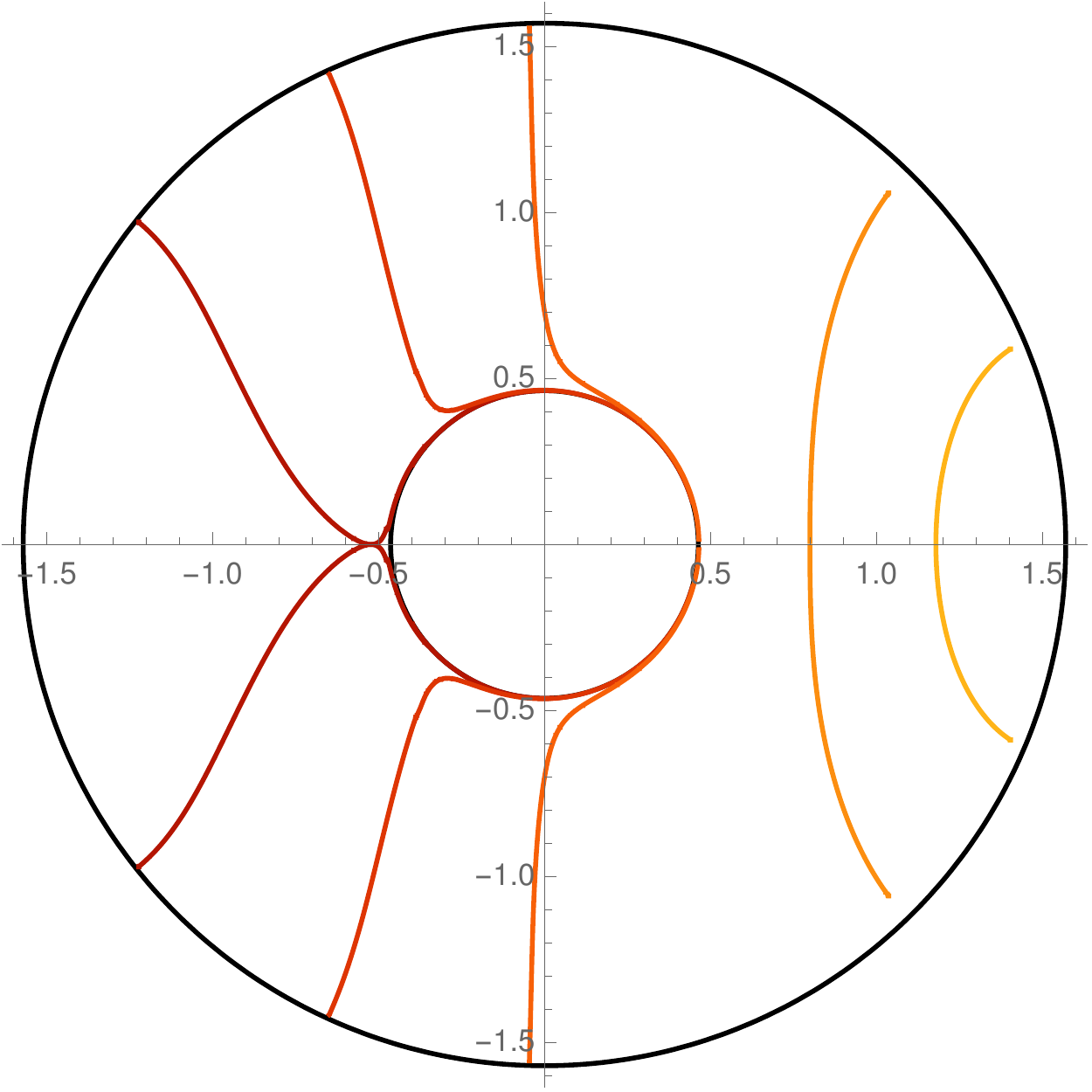}
        \caption{$\lambda<0$}
        \label{fig:fish_lambdapos}
    \end{subfigure}
     \vspace{-2mm}
    \caption{Causal information surfaces $\Xi_\mathcal{A}$ at fixed horizon radius $r_h=0.5$ and fixed Gauss-Bonnet coupling $\lambda$, for several values of $\phi_\mathcal{A}$. $\Xi_\mathcal{A}$ is constructed by shooting the fastest modes with speeds $c_T$ for $\lambda>0$ and $c_S$ for $\lambda<0$ from the lower tip of $D[\mathcal{A}]$ using several values of angular momentum. The darkest red curve corresponds to the critical surface, where the region $\mathcal{A}$ has size $\phi_\mathcal{C}^*$, such that $\Xi_\mathcal{A}$ develops a disconnected component. (a) $\lambda=0.1$, $\phi_\mathcal{C}^*=2.57$. (b) $\lambda=-0.05$, $\phi_\mathcal{C}^*=2.47$.}
    \label{fig:fish}
\end{figure}

Figure \ref{fig:fish} displays the causal information surface for different sizes of the boundary region $\mathcal{A}$ and fixed $\lambda$. For large enough regions, the causal information surface can reach $\phi=\pi$ and the causal wedge develops nontrivial topology. An example of the causal wedge displaying this critical behavior is shown as the darkest red curve in Figure \ref{fig:fish}. We investigate this behavior in more detail in the following.

\subsection{Topological structure of the causal wedge}

Even for simply connected regions $\mathcal{A}$ the causal wedge might display non trivial topology. For example, in a global Schwarzschild-AdS black hole the causal wedge develops a hole for regions with $\phi_\mathcal{A} \geq \phi_\mathcal{C}^*$ for some critical value $\phi_\mathcal{C}^*$, so that the causal holographic information surface has disconneted parts \cite{Hubeny:2013gba}. This change in topology may be a signal of causality violation, since in this situation the causal wedge is more susceptible to intersect the entanglement wedge resulting in a violation of CWI. In Schwarzschild-AdS, it can be explicitly seem that such violation does not happen, because the candidates for the entanglement minimal surface exchange dominance before the causal wedge develops a hole \cite{Hubeny:2013gta}.

The critical points $\phi_\mathcal{C}^*$ can be determined with a good approximation by assuming that the geodesic that cross this critical point has angular momentum $\ell_\text{min}$  \cite{Hubeny:2013gba}, as defined from \eqref{eq:lmin}. Using this assumption, the critical point can be determined as follows. First, at the critical point we have $\phi(r^*)=\pi$, from which we can determine $r_*$. Then, by solving $t(r_*,t_i=-\phi_\mathcal{A})=0$ for $\phi_\mathcal{A}$ we obtain the value of $\phi_\mathcal{C}^*$.

\begin{figure}[t!]
 \centering
 \includegraphics[width=0.7\linewidth]{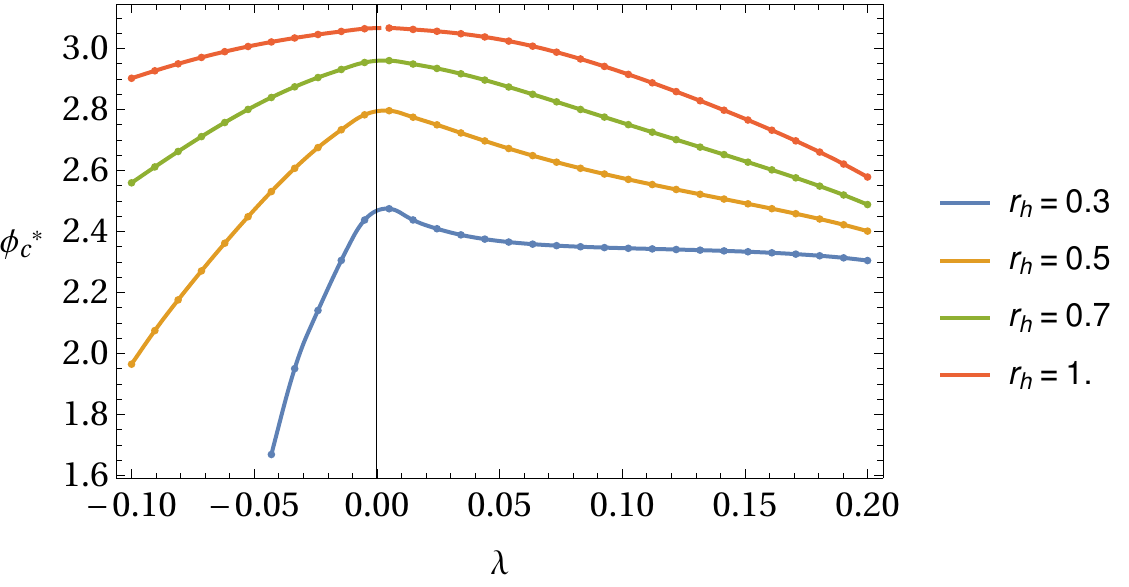}
  \vspace{-5mm}
 \caption{Critical values $\phi_\mathcal{C}^*$, defined such that for $\phi_\mathcal{A}> \phi_\mathcal{C}^*$ the causal wedge displays holes (and the causal information surface is disconneted), for several values of the black hole horizon radius, as a function of the Gauss-Bonnet coupling $\lambda$.}
 \label{fig:phic}
\end{figure}

Figure \ref{fig:phic} shows the values of $\phi_\mathcal{C}^*$ for fixed $r_h$ as a function of $\lambda$. In comparison with the AdS-Schwarzschild case obtained in the $\lambda\to0$ limit, we see that for both positive and negative $\lambda$ the value of $\phi_\mathcal{C}^*$ tends to decrease. By comparing the values of $\phi_\mathcal{C}^*$ where the causal wedge changes topology with the values  $\phi_\mathcal{E}^*$ characterizing the phase transition from Figure \ref{fig:phiE}, we see that for large enough values of $\lambda$, the appearance of a hole in the causal wedge happens before the phase transition, which implies that the entanglement and causal wedges intersect each other and therefore violate CWI. For the values of horizon radius $r_h=0.3,0.5,0.7,1.0$ the upper bounds on $\lambda$ so that   the appearance of holes appear before the transition are, approximately, $\lambda=0.06,0.08,0.10,0.12$. This comparison does not provide lower bounds on $\lambda$.

\subsection{Causal information surface vs. entanglement surface}

In this section we use the causal wedge inclusion (CWI) property (i.e., the property of the causal wedge being contained within the entanglement wedge) to derive constraints on the parameters of the Gauss-Bonnet black hole solution \eqref{eq:metric}. Let us first review why we expect this property to hold by following the argument of \cite{Headrick:2014cta}. The entanglement entropy is not affected by unitary transformations on the reduced density matrix, so it must depend only on the domain $D[\mathcal{A}]$ and not on the particular region $\mathcal{A}$. This implies that the entanglement extremal surface $\Gamma_\mathcal{A}$ cannot be affected by perturbations of the Hamiltonian that have support entirely on $D[\mathcal{A}]$.\footnote{To see this, imagine that we choose a different region $\mathcal{A}'$ such that $D[\mathcal{A}']=D[\mathcal{A}]$ and lying in the past of the perturbation of the Hamiltonian. Since $\mathcal{A}'$ is not affected by the perturbation, $S_{\mathcal{A}'}$ does not change, and since $S_{\mathcal{A}}=S_{\mathcal{A}'}$ we conclude that $\Gamma_\mathcal{A}$ is not affected by the perturbation.} Therefore, $S_{\mathcal{A}}$ has to be causally disconneted from $D[\mathcal{A}]$ and we conclude that $\mathcal{C}(\mathcal{A})\subset\mathcal{E}(\mathcal{A})$.

\begin{figure}[t!]
 \centering
 \includegraphics[width=0.4\linewidth]{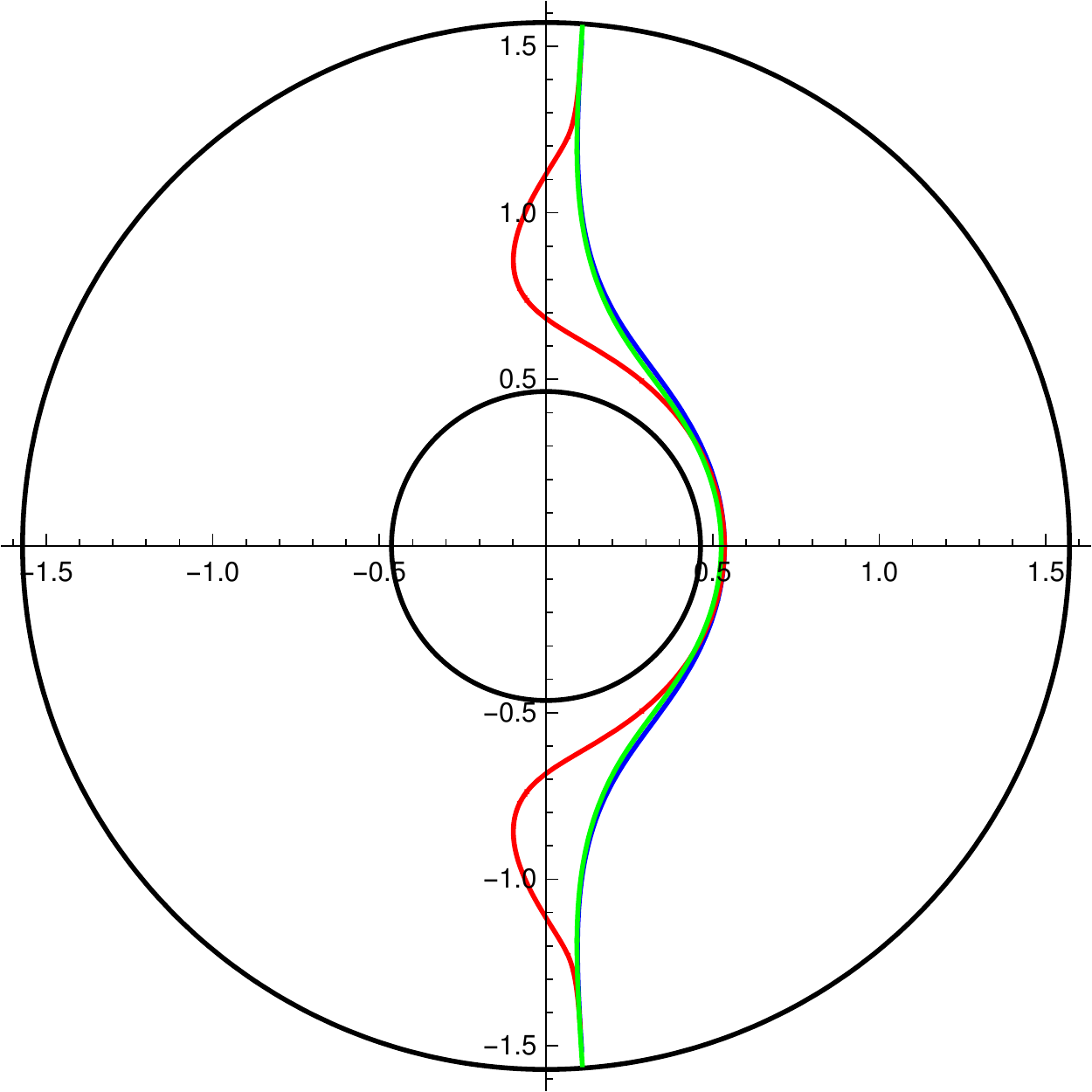}
  \vspace{-2mm}
 \caption{Causal information surface $\Xi_\mathcal{A}$ constructed by shooting the fastest mode of propagation (Red). The equivalent of the causal information surface if one shoots conventional null rays instead of the fastest mode (Blue). Entanglement minimal surface $\Gamma_\mathcal{A}$ (Green). We have set $\lambda=0.2$, $r_h=0.5$, and $\phi_\mathcal{A}=1.5$.}
 \label{fig:wedgenull}
\end{figure}

By looking at the phase transition in entanglement entropy and comparing it to the critical point where the causal information surface becomes disconneted, we have already been able to find a violation of the CWI property for certain values of the coupling $\lambda$. We notice, however, that a violation of CWI can in principle appear for any arbitrary region $\mathcal{A}$, even before the phase transition in the entanglement entropy and before the change in topology of $\mathcal{C}$. Figure \ref{fig:wedgenull} shows an example where CWI is violated. Notice that if we construct the causal wedge via the usual definition by shooting null rays in the physical metric \eqref{eq:metric}, we would not be able to see a violation of CWI.

Here, we would like to systematically compare the causal and entanglement wedges in detail by scanning over all size of spherical boundary regions parametrized by $\phi_\mathcal{A}$ to find the smallest region in the parameters $(r_h,\lambda)$ of the Gauss-Bonnet solution such that the CWI property still holds. Our method to find violations of CWI is to compare the causal information surface $\Xi_\mathcal{A}$ with the entanglement entropy minimal surface $\Gamma_\mathcal{A}$ associated to the same spherical boundary region $\mathcal{A}$. The surface $\Gamma_\mathcal{A}$ is characterized by a curve $\phi_\mathcal{E}(\rho)$, and similarly $\Xi_\mathcal{A}$ is characterized by a curve $\phi_\mathcal{C}(\rho)$. If the difference $\Delta\phi\equiv\phi_\mathcal{E}-\phi_\mathcal{C}$ is less than zero for some $\rho$, then the CWI property is violated. In principle, we need to scan over all possible boundary region sizes parametrized by $\phi_\mathcal{A}$. Numerically, it turns out that the region more susceptible to violation of CWI typically lies in the interval $\frac{\pi}{4}\leq\phi_\mathcal{A}\leq \frac{\pi}{2}$. This behavior is shown in Figure \ref{fig:violation}. In this particular interval the surfaces $\Gamma_\mathcal{A}$ and $\Xi_\mathcal{A}$ are closer to each other and they are deep enough into the bulk so that the effect of the superluminal modes is significant.

\begin{figure}[t!]
    \centering
    \begin{subfigure}[]{0.4\textwidth}
        \includegraphics[width=\textwidth]{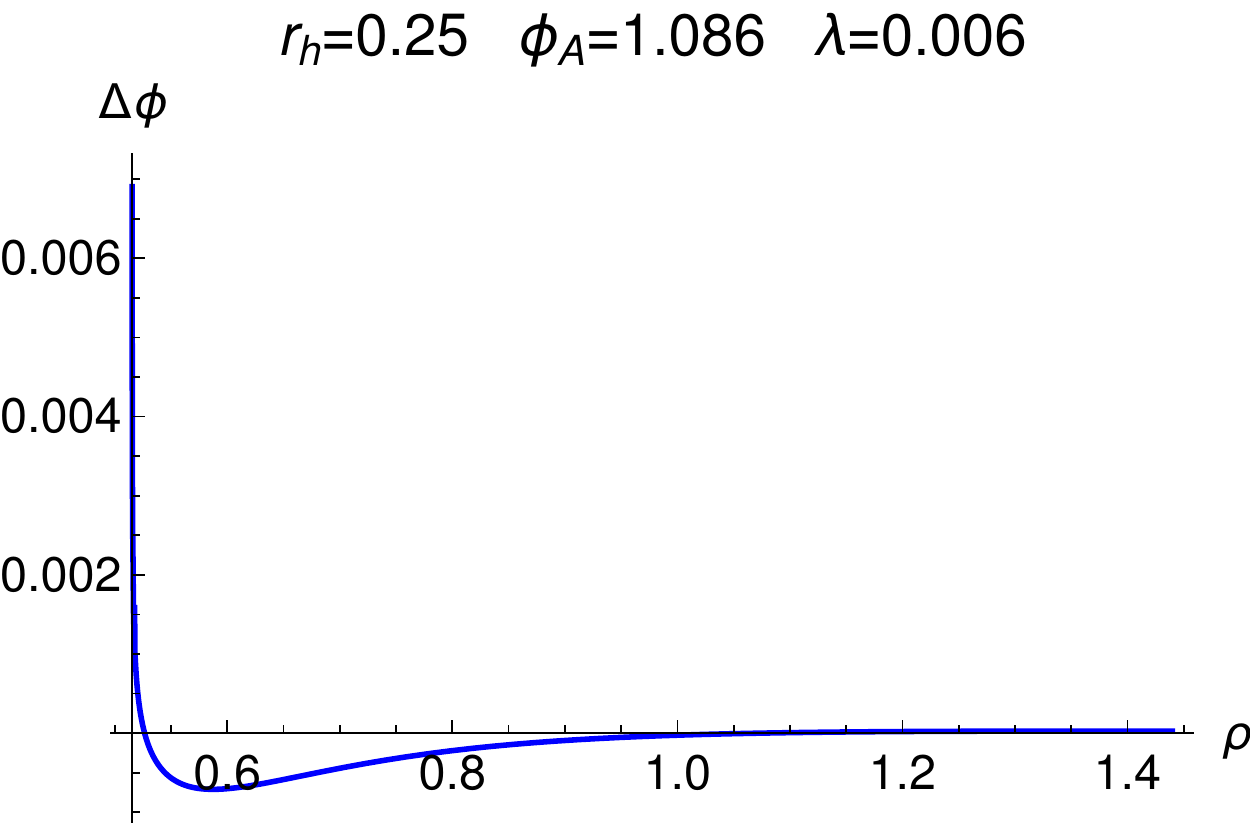}
        \label{fig:violation025}
    \end{subfigure}
    \quad
    \begin{subfigure}[]{0.4\textwidth}
        \includegraphics[width=\textwidth]{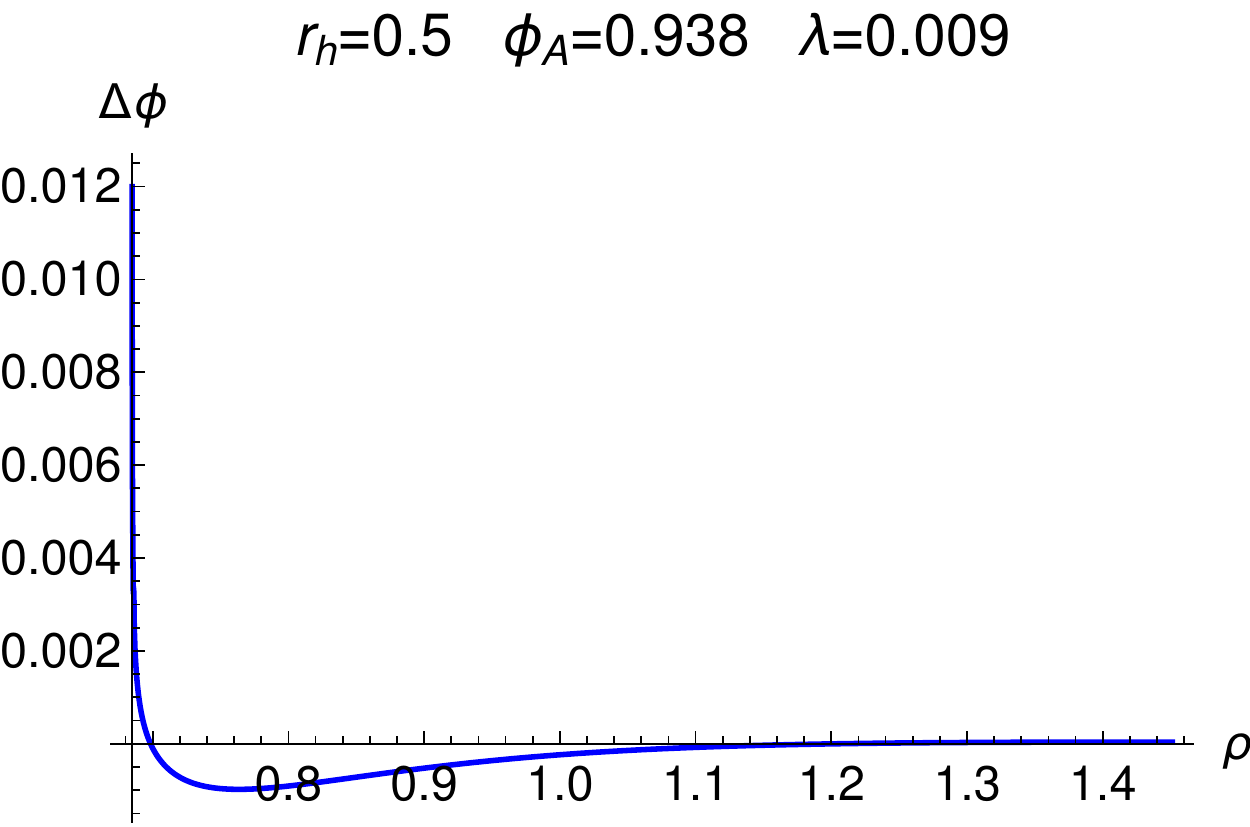}
        \label{fig:violation05}
    \end{subfigure}

    \label{fig:violation}
    \begin{subfigure}[]{0.4\textwidth}
        \includegraphics[width=\textwidth]{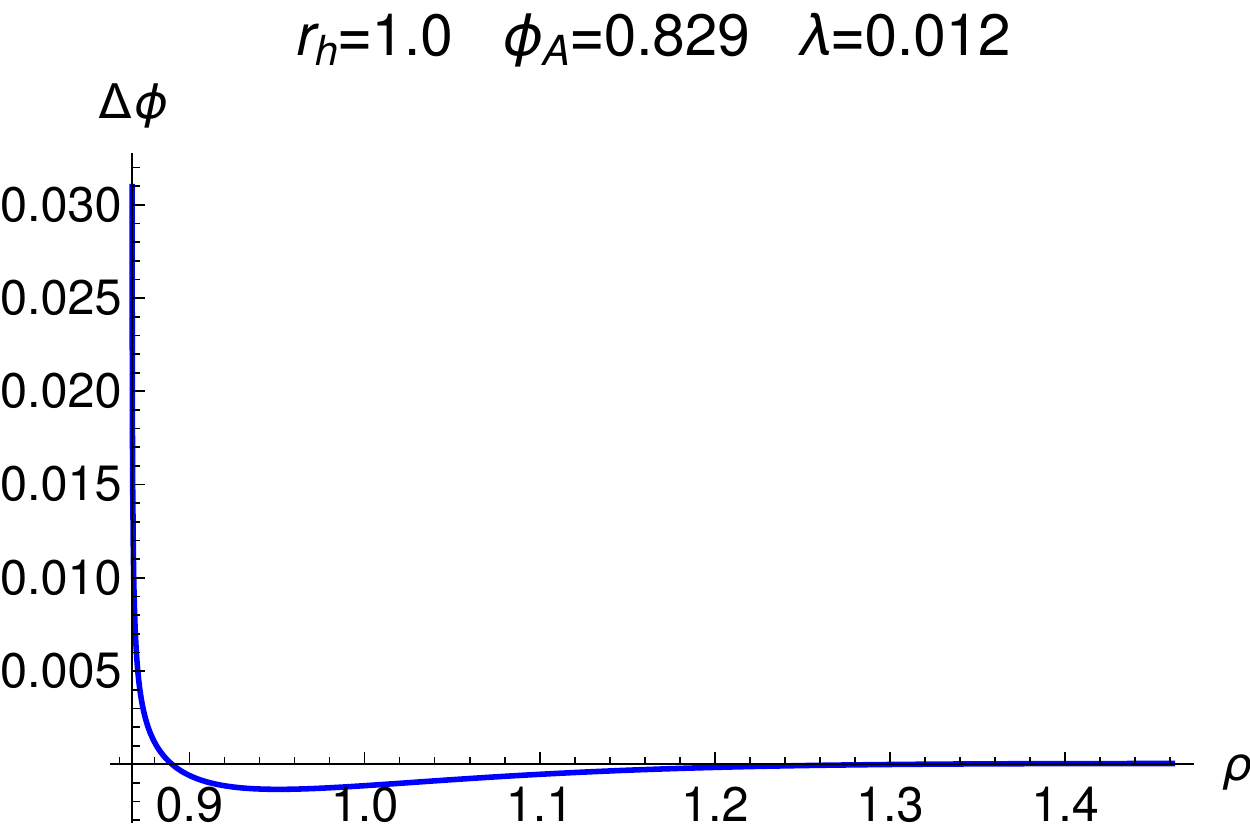}
        \label{fig:violation1}
    \end{subfigure}
    \quad
    \begin{subfigure}[]{0.4\textwidth}
        \includegraphics[width=\textwidth]{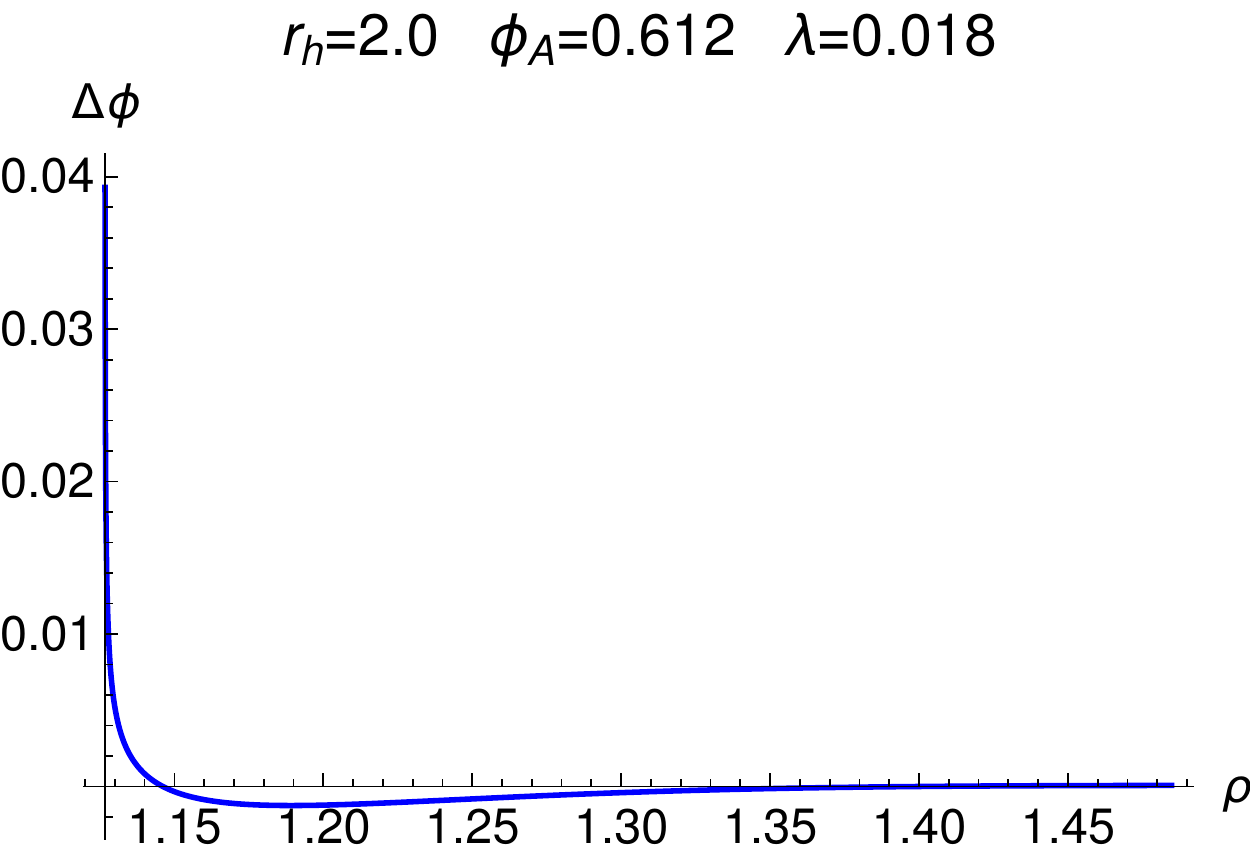}
        \label{fig:violation2}
    \end{subfigure}
     \vspace{-6mm}
    \caption{Angular difference $\Delta\phi=\phi_\mathcal{E}-\phi_\mathcal{C}$ between the entanglement minimal surface $\Gamma_\mathcal{A}$, described by the curve $\phi_\mathcal{E}(\rho)$, and the causal information surface $\Xi_\mathcal{A}$, described by the curve $\phi_\mathcal{C}(\rho)$. The regions where $\Delta\phi<0$ indicate a violation of the CWI property, i.e., for certain choices of parameters $(r_h,\lambda)$ we are able to find a region $\mathcal{A}$ such that $\mathcal{C}(\mathcal{A})\not\subset\mathcal{E}(\mathcal{A})$.}
    \label{fig:violation}
\end{figure}

\begin{figure}[t!]
    \centering
    \begin{subfigure}[]{0.4\textwidth}
        \includegraphics[width=\textwidth]{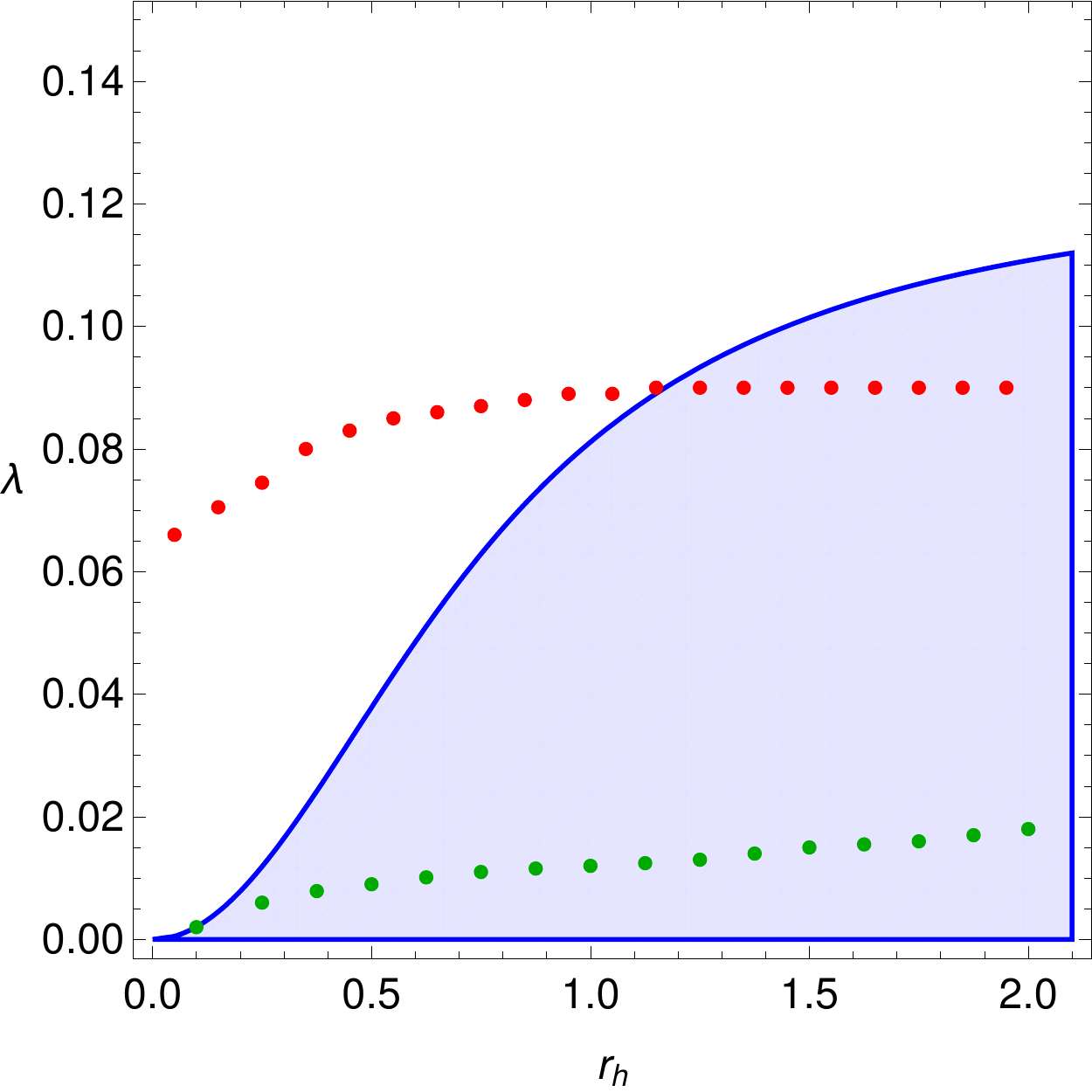}
        \label{fig:constraints_lambdapos}
    \end{subfigure}
    \qquad
    \begin{subfigure}[]{0.4\textwidth}
        \includegraphics[width=\textwidth]{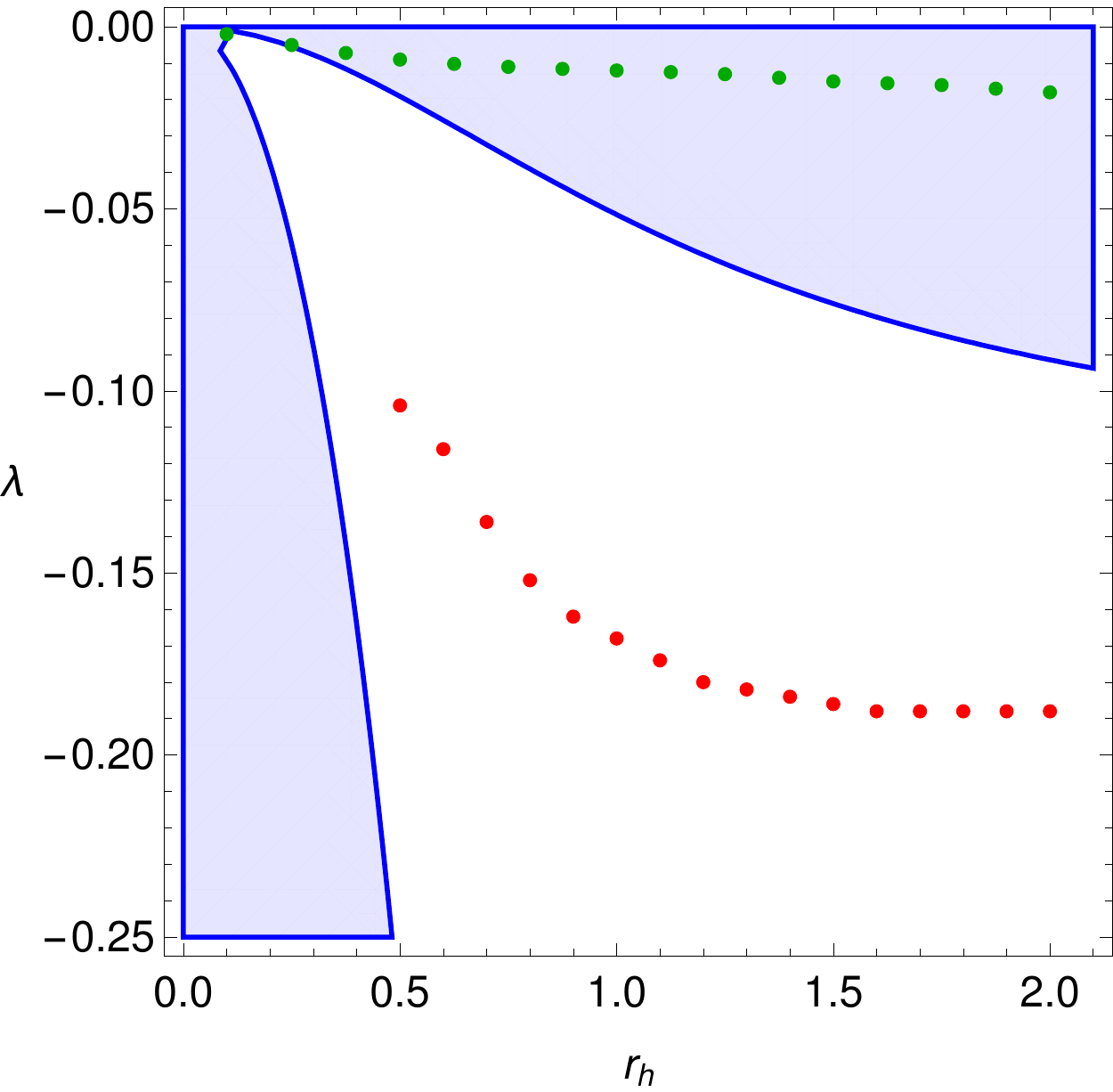}
        \label{fig:constraints_lambdaneg}
    \end{subfigure}
     \vspace{-6mm}
    \caption{
  Constraints on the parameter space $(r_h,\lambda)$ of the spherical AdS Gauss-Bonnet black hole soluton due to: hyperbolicity (Blue), boundary causality \cite{Andrade:2016yzc} (Red), and Causal Wedge Inclusion (Green). The region outside the shaded region or above (below in the right figure) the dots is not allowed.
    }
    \label{fig:constraints}
\end{figure}

Figure \ref{fig:constraints} shows our final result. Above the green dots we have explicitly found a violation of CWI for certain boundary region size. This result shows that in our toy model the implication nBCC $\Rightarrow$ nCWI is obeyed, with the CWI property being much more constraining than boundary causality.

\section{Final remarks\label{sec:5}}

The main observation of this paper was that the construction of the entanglement and causal wedges in general higher order gravities should be performed by shooting geodesics of the fastest mode of the theory. Only in this way, causality in the bulk is properly taken into account in the aforementioned constructs. Assuming that the CWI property must be satisfied by consistency of subregion-subregion duality, this simple but crucial observation translates into a powerful method to constrain higher order derivative theories. In the case of Gauss-Bonnet gravity, our results are summarized in Figure \ref{fig:constraints}. These plots can be interpreted as follows:
\begin{enumerate}
  \item For a given value of $r_h$ (black hole horizon size), we obtain a small window of couplings $\lambda$ for which CWI is satisfied. This is a state-dependent constraint.
  \item Alternatively, we can arrive to a state-independent constraint if we consider all (black hole) states in the theory. In particular, since the window of couplings shrinks to zero size as $r_h\to0$, \emph{any}  $\mathcal{O}(1)$ value of $\lambda$ would lead to an inconsistency in the theory.
\end{enumerate}
The two possibilities here are:  $i)$ we must supplement the gravity action with further (possibly large) higher derivative couplings in such a way that we restore CWI or $ii)$ we can consider Gauss-Bonnet theory on its own, but in that case we can only have perturbative small values of the coupling, of order $\lambda\sim\mathcal{O}(\ell_P^2/L^2)\ll1$. In this scenario, the apparent violations to CWI would be above the UV cutoff $M_P$, so the classical analysis we performed would break down.We point out that a similar conclusion was obtained by Camanho \emph{et.~al.} in \cite{Camanho:2014apa} by  deriving causality constraints from high energy 3-graviton scattering in Gauss-Bonnet gravity.  In \cite{Afkhami-Jeddi:2017rmx} a different approach ---involving a 4-point function and the chaos bound--- was taken to relate the results of  Camanho \emph{et.~al.}  to BCC. Indeed, the relation CWI $\Rightarrow$ BCC  is well established for Einstein gravity (see e.g., \cite{Akers:2016ugt}). Our results imply  that  this arrow  also holds true for Gauss-Bonnet gravity, which can easily be checked by our final plots in Figure \ref{fig:constraints}.

Let us end with a couple of open questions and extensions  that are worth exploring.
\vspace{-2mm}

\paragraph{Further constrains from subregion duality:}
Entanglement Wedge Nesting (EWN) is the property that $\mathcal{D}[\mathcal{A}]\subset\mathcal{D}[\mathcal{B}]$ implies $\mathcal{E}(\mathcal{A})\subset\mathcal{E}(\mathcal{B})$. This property is essential in subregion duality, concretely, for the reconstruction of the entanglement wedge. In Einstein gravity, it can been shown that EWN $\Rightarrow$ CWI $\Rightarrow$ BCC (see e.g., \cite{Akers:2016ugt}), but the proof does not hold for higher order gravities. In Gauss-Bonnet gravity, the extremal surfaces obey the inclusion property. Therefore, if there is a violation of EWN one should look at the entire entanglement wedge, constructed by shooting geodesics of the mode with the fastest propagation. Since the surfaces $\Gamma_{\mathcal{A}}$ are nested, there is little room for its violation but it might still happen due to caustics in the entanglement wedge. In that case, EWN might give even stronger constraints than CWI and could  be used to close the window of allowed couplings even further. Another interesting possibility would be to consider quantum corrections to both CWI and EWN. This could be done using quantum extremal surfaces \cite{Engelhardt:2014gca,Engelhardt:2019hmr}, although it might be difficult to implement in practice.
\vspace{-2mm}

\paragraph{Extension beyond Gauss-Bonnet:}

Our method to derive constraints based on CWI can in principle be applied to any higher order gravity theory. In order to do so one would need two basic ingredients. The first one is to determine whether superluminal propagation exists by studying the different gravitational modes. The second one is the construction of the entanglement and causal wedges. As explained in the paper, this must be done by shooting the fastest mode of propagation instead of null modes. Finally, having the two wedges one could easily test the inclusion property for the theory in consideration. While this might be technically too involved in a general higher derivative theories, there are cases where the problem is tractable. For example, the hyperbolicity of Lovelock and Horndeski gravity was investigated in \cite{Reall:2014pwa,Papallo:2017qvl,Minamitsuji:2015nca}, the entanglement functional in \cite{Dong:2013qoa,Caceres:2017lbr} and the fluctuations in \cite{Kobayashi:2014wsa}.

\paragraph{Universality in higher order gravities:}
One natural question is the universality of our results in generic higher order gravities. Does the qualitative behavior of Figure \ref{fig:constraints} persists for general quadratic and cubic higher order gravities? Or can there be gravity theories for which the higher derivative couplings do not need to be taken perturbatively small? Is this result general for theories with superluminal propagation in the bulk? And what happens for higher order gravity theories with \emph{no} superluminal propagation?
Can we still constrain these theories with CWI, or some other property of subregion duality?
\vspace{-2mm}

\paragraph{Gao-Wald theorem:}
The Gao-Wald theorem \cite{Gao:2000ga} addresses the question of bulk spacetime
being compatible with boundary causality, in the sense that we cannot have
faster propagation between two boundary points through a path in the bulk. The theorem holds in general for any asymptotically AdS spacetime in Einstein gravity, and states that the Averaged Null Energy Condition (ANEC) is a sufficient condition to ensure that boundary causality is obeyed. If we do not assume Einstein gravity, the ANEC is replaced by the Null Curvature Condition (ANCC).
The converse of the Gao-Wald theorem is not true in general, i.e., preserving the causal structure at the boundary does not imply the ANCC to be satisfied in the bulk. Progress to find the necessary and sufficient conditions to preserve boundary causality have been made by Engelhardt and Fischetti \cite{Engelhardt:2016aoo}, where they found the necessary and sufficient conditions at the linear level when the spacetime is a perturbation of pure AdS. However, both Gao-Wald and Engelhardt-Fischetti implicitly assume that the causal structure is controlled by null rays, so they do not apply for theories with superluminal propagation. It would be interesting to see whether these theorems can be modified to take the superluminal modes into account.
\vspace{-2mm}

\paragraph{Holographic complexity:}
Similarly to the modifications in the definitions of the causal and entanglement wedges due to the existence of bulk superluminal propagation, we should also be careful when considering other bulk regions defined via a causal domain, such as the WdW patch used to compute holographic complexity \cite{Brown:2015bva,Brown:2015lvg}. In  Gauss-Bonnet, the superluminal modes travel along the angular directions, while gravity still travels at the speed of light along a radial path. However, we do not have to construct the WdW patch using purely radial geodesics, we can use geodesics with angular momentum. Consequently, the WdW patch could be modified by the presence of superluminal modes.  Also,  subregion complexity-action \cite{Carmi:2016wjl} involves the entanglement wedge, so we expect  superluminal modes to have a clear effect in  its behavior. We point out that previous investigations of complexity in Gauss-Bonnet gravity include \cite{An:2018dbz, Cano:2018aqi}, but they do not consider the effects of these modes. It would be interesting to  investigate complexity-action in the region of parameters where we know that boundary causality and CWI are satisfied or violated. It would be also interesting to investigate if the relation between complexity and entanglement found in \cite{Agon:2018zso,Caceres:2018luq} persists in higher derivative theories.

\vspace{1mm}
We hope to come back to some of these points in the near future.

\section*{Acknowledgements}

We would like to thank Cesar Ag\'on, Ben Freivogel, Cindy Keeler, Sagar Lokhande and  Manus Visser for useful discussions and comments on the manuscript. EC and ASM are supported by the National Science Foundation (NSF) Grants No. PHY-1620610 and No. PHY-1820712. EC would also like to thank the Centro de Ciencias de Benasque Pedro Pascual as well as the organizers of the workshop ``Gravity - New perspectives from strings and higher dimensions'' for the welcoming atmosphere and hospitality during the final stages of this project.
JFP is supported by the Netherlands Organization for Scientific Research (NWO) under the VENI grant No. 680-47-456/1486.

\appendix

\section{Entanglement entropy functional in Gauss-Bonnet} \label{sec:functional}

Let us specialize the functional \eqref{eq:functional} for the black hole geometry
\eqref{eq:metric}. This is essentially the same computation of \cite{Sun:2016til}. Using the spherical symmetry, we can parametrize the surface $\Gamma_{\mathcal{A}}$ by $(r(\sigma),\phi(\sigma))$. The induced metric on this hypersurface is
\begin{align}
h_{ab}dx^adx^b&=\left(\frac{{r'}^2(\sigma)}{f(r(\sigma))}+r^2(\sigma){\phi'}^2(\sigma)\right)d\sigma^2+r^2(\sigma)\sin^2(\phi(\sigma))d\Omega^2_2\nonumber\\
&\equiv h_{\sigma\sigma}(\sigma)d\sigma^2+e^{2F(\sigma)}d\Omega^2_2,
\label{eq:indmetric}
\end{align}
where the indices $a,b$ run over $(\sigma,\chi,\omega)$. The scalar curvature on $\Gamma_{\mathcal{A}}$ is
evaluated to
\begin{equation}
\mathcal{R}
=2e^{-2F}-\frac{6{F'}^2}{h_{\sigma\sigma}}+\frac{2F'h_{\sigma\sigma}'}{h^2_{\sigma\sigma}}-\frac{
	4F''}{h_{\sigma\sigma}}.
\end{equation}
The integral over $\Gamma_{\mathcal{A}}$ in \eqref{eq:functional} becomes (after integrating over the 2-sphere)
\begin{align}
&\frac{1}{4G_N}\int_{\Gamma_{\mathcal{A}}}\!\!\! d^{3}x\sqrt{h}(1+\lambda L^2 \mathcal{R})=
\nonumber\\
&\qquad=\frac{\pi}{G_N}\int d\sigma\sqrt{h_{\sigma\sigma}}(e^{2F}+2\lambda L^2)+\frac{2\pi\lambda
	L^2}{G_N}\int d\sigma
\frac{e^{2F}}{\sqrt{h_{\sigma\sigma}}}\left(-3{F'}^2+\frac{F'h_{\sigma\sigma}'}{h_{\sigma\sigma}}
-2F''\right)\nonumber\\
&\qquad= \frac{\pi}{G_N}\int
d\sigma\left(\sqrt{h_{\sigma\sigma}}(e^{2F}+2\lambda L^2)+ 2\lambda
L^2\frac{e^{2F}{F'}^2}{\sqrt{h_{\sigma\sigma}}}\right) -\frac{4\pi\lambda
	L^2}{G_N}\frac{e^{2F}F'}{\sqrt{h_{\sigma\sigma}}}\Big|_{\sigma=\sigma_{bdy}}.
\label{eq:bdyterm}
\end{align}
The boundary term is expected to cancel the extrinsic curvature term in \eqref{eq:functional}. As
a check, we can compute the extrinsic curvature by defining the outward unit vector normal  to
$\Gamma_{\mathcal{A}}$
\begin{equation}
n_a=\sqrt{h_{\sigma\sigma}}\delta_{\sigma a},\quad n^a=\sqrt{h^{\sigma\sigma}}\delta_{\sigma a}.
\end{equation}
The extrinsic curvature evaluated at $\partial \Gamma_{\mathcal{A}}$ is
\begin{equation}
\mathcal{K}
=h^{ab}\nabla_an_b\Big|_{\sigma=\sigma_{bdy}}=\frac{2F'}{\sqrt{h_{\sigma\sigma}}}\Big|_{
	\sigma=\sigma_{bdy}}.
\end{equation}
The integral of the extrinsic curvature in \eqref{eq:functional} becomes (using $\sqrt{\tilde{h}}=e^{2F}$)
\begin{equation}
\frac{\lambda L^2}{2G_N}\int_{\partial\Gamma_{\mathcal{A}}}\!\!\! d^2x\sqrt{\tilde{h}}\mathcal{K}=\frac{4\pi\lambda
	L^2}{G_N}\frac{e^{2F}F'}{\sqrt{h_{\sigma\sigma}}}\Big|_{\sigma=\sigma_{bdy}},
\end{equation}
which precisely cancels the boundary term in \eqref{eq:bdyterm}. The functional
\eqref{eq:functional} is then
\begin{equation}
S_\mathcal{A}=\frac{\pi}{G_N}\int d\sigma\left(\sqrt{h_{\sigma\sigma}}(e^{2F}+2\lambda
L^2)+\frac{2\lambda L^2e^{2F}{F'}^2}{\sqrt{h_{\sigma\sigma}}}\right).
\end{equation}
Using the explicit form of $F$ from the induced metric \eqref{eq:indmetric} we obtain
\begin{equation}
S_\mathcal{A}=\frac{\pi}{G_N}\int d\sigma\left[\sqrt{h_{\sigma\sigma}}\left(r^2\sin^2\phi+2\lambda
L^2\right)+\frac{2\lambda L^2}{\sqrt{h_{\sigma\sigma}}}\left(r\phi'\cos\phi +
r'\sin\phi\right)^2\right]
\end{equation}
where $r=r(\sigma),\phi=\phi(\sigma)$, the prime denotes derivative w.r.t. $\sigma$, and
\begin{equation}
h_{\sigma\sigma}(\sigma)=\frac{{r'}^2(\sigma)}{f(r(\sigma))}+r^2(\sigma){\phi'}^2(\sigma).
\end{equation}

\end{document}